\documentclass[showpacs,preprintnumbers,amsmath,amssymb, nofootinbib]{revtex4}
\usepackage{fontenc}
\usepackage{graphicx}

\usepackage[dvips]{hyperref}

\usepackage{amsmath}

\begin{document}

\title{Chern-Simons black holes: scalar perturbations, mass  and area spectrum \\ and greybody factors
}
\author{Pablo Gonz\'{a}lez}
\email{pablo.gonzalezm@mail.udp.cl} \affiliation{Instituto de
F\'{\i}sica, Pontificia Universidad Cat\'olica de Valpara\'{\i}so,
Casilla 4950, Valpara\'{\i}so, Chile} \affiliation{Universidad
Diego Portales, Casilla 298-V, Santiago, Chile.}
\author{Eleftherios Papantonopoulos}
\email{lpapa@central.ntua.gr} \affiliation{
 Department of Physics, National Technical University of
Athens, Zografou Campus GR 157 73, Athens, Greece. }
\author{Joel Saavedra}
\email{joel.saavedra@ucv.cl} \affiliation{Instituto de
F\'{\i}sica, Pontificia Universidad Cat\'olica de Valpara\'{\i}so,
Casilla 4950, Valpara\'{\i}so, Chile.}
\date{\today}

\begin{abstract}

We study the Chern-Simons black holes in $d-$dimensions and we
calculate analytically the quasi-normal modes of the scalar
perturbations and we show that they depend on the highest power of
curvature present in the Chern-Simons theory. We obtain the mass
and area spectrum  of these black holes and we show that they have
a strong dependence on the topology of the transverse space and
they are not evenly spaced. We also calculate analytically the
reflection and transmission coefficients and the absorption cross
section and we show that at low frequency limit there is a range
of modes which contributes to the absorption cross section.

\end{abstract}

\maketitle

\section{Introduction}

Chern-Simons black holes are special solutions of gravity theories
in higher than four dimensions  which contain higher powers of
curvature. These theories are consistent Lanczos-Lovelock theories
resulting in second order field equations for the metric with well
defined AdS asymptotic solutions. For spherically symmetric
topologies, these black holes are labelled by an integer $k$ which
specifies the higher order of curvature present in the
Lanczos-Lovelock action and it is related to the dimensionality
$d$ of spacetime by the relation
$d-2k=1$~\cite{Crisostomo:2000bb}. These solutions where further
generalized to flat and hyperbolic topologies~\cite{Aros:2000ij}.

For spherical topologies the Chern-Simons black holes have similar
causal structure as the (2+1)-dimensional BTZ black
hole~\cite{Banados:1992wn}, and they have positive specific heat
and therefore thermodynamical stability. For hyperbolic
topologies, the Chern-Simons black holes resemble to the
topological black holes~\cite{topological} in their zero mass
limit, and their thermodynamic behaviour was studied
in~\cite{Aros:2000ij}.

An illustrative  example of Chern-Simons black holes is provided
by the Gauss-Bonnet theory for $d=5$ and $k=2$. Static local
solutions of this theory are well studied over the
years~\cite{gbpapers}. This theory has two branches of solutions.
If there is a fine tuning between $k$ and the Gauss-Bonnet
coupling constant $\alpha$, the two solutions coincide to the
Chern-Simons black hole solution which has maximum symmetry. This
is known as the Chern-Simons limit (for a review
see~\cite{Charmousis:2008kc}). The stability of these solutions
has also been studied~\cite{stability}. It was found
in~\cite{Charmousis:2008ce} that one of these solutions suffers
from ghost-like instability up to the strongly coupled
Chern-Simons limit where linear perturbation theory breaks down.
Therefore, to study the stability at the exact Chern-Simons limit
we have to go beyond perturbation theory. It was speculated
 in~\cite{Charmousis:2008ce} that Chern-Simons black holes could be a
transitional configuration between  the two branches of solutions.

In this work we will consider  a matter distribution outside the
horizon of a Chern-Simons black hole with hyperbolic topology
parameterized by a scalar field. We will perturb the scalar field
assuming that there is no back reaction on the metric. This  will
result in the calculation of the quasi-normal modes (QNMs) which
are characterized by a spectrum that is independent of the initial
conditions of the perturbation and depends only on the black hole
parameters and on the fundamental constants of the system (for a
recent review see~\cite{Siopsis:2008xz}). A novel and interesting
feature of our calculation is that the QNMs of the Chern-Simons
black holes depend also on the curvature parameter $k$. As we will
discuss, because of the presence of the curvature parameter $k$ as
we approach the Chern-Simons limit the curvature singularity
becomes milder. This may give detectable signature through the
QNMs in the gravitational waves experiments.

Another interesting feature of the QNMs is their connection to
thermal conformal field theories. According to the AdS/CFT
correspondence~\cite{Maldacena:1997re}, classical gravity
backgrounds in AdS space are dual to conformal field theories at
the boundary (for a review see~\cite{Aharony:1999ti}). Using this
principle it was established~\cite{Horowitz:1999jd} that the
relaxation time of a thermal state of the boundary thermal theory
is proportional to the inverse of the imaginary part of the QNMs
of the dual gravity background. Therefore, the knowledge of the
QNMs spectrum determines how fast a thermal state in the boundary
theory will reach thermal equilibrium. We will show that, the rate
at which a scalar field in the background of a Chern-Simons black
hole will decay or the rate the boundary thermal theory will reach
thermal equilibrium,  depends on the value of the curvature
parameter $k$.

The QNMs were also studied in relation to the quantum area
spectrum of the black hole horizon. Bekenstein
\cite{Bekenstein:1974jk} was the first who proposed the idea that
in quantum gravity the area of black hole horizon is quantized
leading to a discrete spectrum which is evenly spaced. An
interesting proposal was made by Hod~\cite{Hod:1998vk} who
conjectured that the asymptotic QNM frequency is related to the
quantized black hole area. The black hole spectrum can be obtained
imposing the Bohr-Sommerfeld quantization condition to an
adiabatic invariant quantity involving the energy $E$ and the
vibrational frequency $\omega(E)$~\cite{Kunstatter:2002pj}.
Identifying $\omega(E)$ with the real part $\omega_R$ of the QNMs,
the Hod's conjecture leads to an expression of the quantized black
hole area, which however is not universal for all black hole
backgrounds. Furthermore it was argued~\cite{Maggiore:2007nq},
that in the large damping limit the identification of $\omega(E)$
with the imaginary part of the QNMs could lead to the Bekenstein
 universal bound~\cite{Bekenstein:1974jk}.

Both approaches have been followed  in the
literature~\cite{areaquantization}.  In this work we will
calculate the area spectrum and the mass spectrum of a
Chern-Simons black hole with hyperbolic topology. We will show
that there is a strong dependence of the spectrums on the
hyperbolic geometry and that they are not evenly spaced.

The knowledge of black holes perturbations is also useful in
studying the Hawking radiation. The Hawking radiation is a
semiclassical effect and it gives the thermal radiation emitted by
a black hole. However, black holes do not radiate strictly
blackbody type radiation due to the well known frequency dependent
greybody factors.  These factors arise from frequency-dependent
potential barriers outside the horizon which filter the initially
blackbody spectrum emanating from the horizon
\cite{Maldacena:1996ix}. In asymptotically AdS spacetimes, the
light rays can reach spatial infinity and return to the origin in
finite time. Black holes in this kind of spacetimes are in thermal
equilibrium with their environment. So, the radiation produced at
black hole horizon is all re-absorbed and the radiation which
reaches spatial infinity is reflected back, with part of this
radiation travelling all the way through to the black hole
horizon, and the rest being reflected back to spatial infinity
\cite{Harmark:2007jy}.

In the present work, the reflection and the transmission
coefficients and the greybody factors of Chern-Simons black holes
   are computed analytically. We make a
numerical analysis of their behaviour in five-dimensions and in
the low frequency limit and we found that contrary to the
Schwarzschild-AdS case there is a range of modes with high
angular momentum which contributes to the
absorption cross section.

The paper is organized as follows. In Sec. II we give a brief
review of the Chern-Simons theory. In Sec. III we calculate the
exact QNMs of the scalar perturbations  of d-dimensional
Chern-Simons black holes with hyperbolic topology. In Sec. IV
using the calculated QNMs, the mass and area spectrum of
Chern-Simons black holes  are presented. In Sec. V we calculate
the  reflection and the transmission coefficients and the greybody
factors of Chern-Simons black holes. Finally, our conclusions are
in Sec. VI.

\section{The Chern-Simons Theory}

The Einstein tensor is the only symmetric and conserved tensor
depending on the metric and its derivatives, which is linear in
the second derivatives of the metric. The field equations arise
from the  Einstein-Hilbert action  with a cosmological constant
$\Lambda$. In higher dimensions, the potential problem is to find
the most general action that gives rise to a set of second order
field equations. The solution to this problems is the
Lanczos-Lovelock (LL) action \cite{LL}. This action is non linear
in the Riemann tensor and differs from the Einstein-Hilbert action
only if the spacetime has more than $4$ dimensions. Therefore, the
Lanczos-Lovelock action is the most natural extension of general
relativity in higher dimensional spacetimes. In  $d-$dimensions it
can be written as follow
\begin{equation}\label{dtopologicalaction}\
I_{k}=\kappa \int \sum_{q=0}^{k}c_{q}^{k}L^{q},
\end{equation}%
with
\begin{equation}\label{polinomio}\
L^{q}=\epsilon_{\alpha_{1}...\alpha_{d}}R^{\alpha_{1}\alpha_{2}}...R^{\alpha_{2q-1}\alpha_{2q}}e^{\alpha_{2q+1}}...e^{\alpha_{d}},
\end{equation}%
where $e^{\alpha}$ and $R^{\alpha\beta}$ stand for the vielbein
and the curvature two-form respectively and $\kappa$ and $l$ are
related to the gravitational constant $G_{k}$ and the cosmological
constant $\Lambda$ through
\begin{equation}\label{definitionk}\
\kappa=\frac{1}{2(d-2)!\Omega _{d-2}G_{k}},
\end{equation}%
\begin{equation}\label{lambda}\
\Lambda=-\frac{(d-1)(d-2)}{2l^2},
\end{equation}%
and $\alpha_q:=c_{q}^{k}$ where
$c_{q}^{k}=\frac{l^{2(q-k)}}{d-2q}(^{k}_{q})$ for $q\leq k$ and
vanishes for $q>k$, with $1\leq k\leq [\frac{d-1}{2}]$ ($[x]$
denotes the integer part of $x$) and $\Omega _{d-2}$ corresponds
to the volume of a unit $(d-2)$-dimensional sphere. Static black
hole-like geometries with spherical topology were
found~\cite{Crisostomo:2000bb} possessing topologically nontrivial
AdS asymptotics.  These theories and their corresponding solutions
were classified by an integer $k$, which corresponds to the
highest power of curvature in the Lagrangian. If $d-2k=1$, the
solutions are known as Chern-Simons black holes (for a review on
the Chern-Simons theories see \cite{Zanelli:2005sa}). These
solutions were further generalized to other
topologies~\cite{Aros:2000ij} and they can be described in general
by a non-trivial transverse spatial section $\sum_{\gamma}$ of
$(d-2)$-dimensions  labelled by the constant $\gamma=+1, -1, 0$
that represents the curvature of the transverse section,
corresponding to a spherical, hyperbolic or plane section
respectively. The solution describing a black hole in a free
torsion theory can be written as~\cite{Aros:2000ij}
\begin{equation}\label{dtopologicalmetric}\
ds^{2}=-\Big{(} \gamma +\frac{r^{2}}{l^{2}}-\alpha \Big{(} \frac{2G_{k}\mu }{%
r^{d-2k-1}}\Big{)} ^{\frac{1}{k}}\Big{)} dt^{2}+\frac{dr^{2}}{\Big{(} \gamma +%
\frac{r^{2}}{l^{2}}-\alpha \Big{(} \frac{2G_{k}\mu }{r^{d-2k-1}}\Big{)} ^{%
\frac{1}{k}}\Big{)} }+r^{2}d\sigma _{\gamma }^{2}~,
\end{equation}%
where $\alpha =(\pm 1)^{k+1}$ and the constant $\mu$ is related to
the horizon $r_{+}$ through
\begin{equation}\label{relationr}\
\mu =\frac{r_{+}^{d-2k-1}}{2G_{k}}(\gamma
+\frac{r_{+}^{2}}{l^{2}})^{k}~,
\end{equation}%
and to the mass $M$ by
\begin{equation}\label{relationM}\
\mu =\frac{\Omega _{d-2}}{\Sigma _{d-2}}M+\frac{1}{2G_{k}}\delta
_{d-2k,\gamma }~,
\end{equation}%
here $\Sigma _{d-2}$ denotes the volume of the transverse space.
As can be seen in (\ref{dtopologicalmetric}), if $d-2k\neq1$ the
$k$ root makes the curvature singularity milder than the
corresponding black hole of the same mass. At the exact
Chern-Simons limit $d-2k=1$, the solution has similar structure
like the (2+1)-dimensional BTZ black hole with a string-like
singularity.

We are merely interested here for the hyperbolic topology with
$\gamma=-1$. The reason is that Chern-Simons theories with
$\gamma=-1$ always possess a single horizon and have interesting
thermodynamic behaviour for even and odd $k$ \cite{Aros:2000ij}.
In this case $d\sigma^{2}_{-1}$ in (\ref{dtopologicalmetric}) is
the line element of the $(d-2)$-dimensional manifold $\sum_{-1}$,
which is locally isomorphic to the hyperbolic manifold $H^{d-2}$
or pseudosphere, which is a non-compact ($d-2$)-dimensional space
of constant negative curvature and of the form
\begin{equation}\label{eqSigma}
\sum=H^{d-2}/\Gamma \quad \textrm{,\quad  $\Gamma\subset
O(d-2,d-1)$}~,
\end{equation}
where $\Gamma$ is a freely acting discrete subgroup (i.e., without
fixed points) of isometries.  This space becomes a compact space
of constant negative curvature with genus $g \ge 2$ by
identifying, according to the connection rules of the discrete
subgroup $\Gamma$, the opposite edges of a $4g$-sided polygon
whose sides are geodesics and is centered at the origin
 of the pseudosphere~\cite{topological,
Balazs:1986uj}. In the case of $d=2$, an octagon is the simplest
such polygon, yielding a compact surface of genus $g=2$ under
these identifications. Thus in general, the $(d-2)$-dimensional
manifold $\sum_{-1}$ is a compact Riemann $(d-2)$-surface of genus
$g \geq2$. The configuration (\ref{dtopologicalmetric}) is an
asymptotically locally AdS spacetime.

\section{$d-$ dimensional Quasi-normal modes of scalar perturbations}

To obtain an exact expression for the quasi-normal modes of scalar
perturbations of a Chern-Simons black hole in $d-$dimensions we
need to impose boundary conditions on asymptotically AdS
spacetime. These are that at the horizon there exist only ingoing
waves and  the vanishing of the flux of the field at infinity. The
metric of Chern-Simons theories is
\begin{equation}\label{metric2}\
ds^{2}=-f(r)dt^{2}+\frac{1}{f(r)}dr^{2}+r^{2}d\sigma
_{\gamma}^{2}~,
\end{equation}
where,
\begin{equation}
f(r)=\gamma +\frac{r^{2}}{l^{2}}-\alpha(2\mu
G_{k})^{\frac{1}{k}}~,
\end{equation}
and the horizon is located at
\begin{equation}\label{horizon}\
r_{+}=l\sqrt{\alpha(2\mu G_{k})^{\frac{1}{k}}-\gamma}~.
\end{equation}
Considering the horizon geometry with a negative curvature
constant, $\gamma=-1$, the allowed range of $\mu$ for $r_+\geq0$
are: if $k$ is odd, $\alpha=1$, $\mu\geq\frac{-1}{2G_k}$; if $k$
is even and $\alpha=1$, $\mu \geq0$ and if $k$ is even and
$\alpha=-1$, $\frac{1}{2G_k}\geq\mu\geq0$ \cite{Aros:2000ij}.
Performing the change of variables $z=p-l^2/r^2$, where
$p=1+\alpha(2\mu G_{k})^{\frac{1}{k}}$ and $t=lt$, the metric
 (\ref{metric2}) can be written
\begin{equation}\label{metric3}\
ds^{2}=-f(z)l^2dt^{2}+\frac{l^2}{4f(z)(p-z)^3}dz^{2}+\frac{l^{2}}{p-z}d\sigma
_{\gamma}^{2}~,
\end{equation}
where
\begin{equation}
f(z)=\frac{1-p^2(1-\frac{z}{p})}{p(1-\frac{z}{p})}~.
\end{equation}
The horizon  (\ref{horizon}) now is located at $r_{+}=l\sqrt{p}$.
With the definition $x=1-\frac{z}{p}$ the metric (\ref{metric3})
becomes
\begin{equation}\label{metric4}\
ds^{2}=-\frac{(1-p^2x)l^2}{px}dt^{2}+\frac{l^2}{4(1-p^2x)x^2}dx^{2}+\frac{l^{2}}{px}d\sigma
_{\gamma}^{2}~.
\end{equation}
If we define  $y=p^2x$  the metric (\ref{metric4}) can be written
as
\begin{equation}\label{metric5}\
ds^{2}=-\frac{(1-y)l^2
p}{y}dt^{2}+\frac{l^2}{4(1-y)y^2}dy^{2}+\frac{l^{2}p}{y}d\sigma
_{\gamma}^{2}~,
\end{equation}
and introducing $v=1-y$ we finally obtain
\begin{equation}\label{finalmetric}\
ds^{2}=-\frac{l^2 p
v}{(1-v)}dt^{2}+\frac{l^2}{4v(1-v)^2}dv^{2}+\frac{l^{2}p}{1-v}d\sigma
_{\gamma}^{2}~.
\end{equation}

A minimally coupled scalar field to curvature in the background of
a Chern-Simons black hole in $d-$dimensions is given by the
Klein-Gordon equation
\begin{equation}\label{waveequation}\
\nabla^2\varphi=m^2\varphi~.
\end{equation}
We adopt the ansatz $\varphi=R(v)Y(\sum)e^{-i\omega t}$, where $Y$
is a normalizable harmonic function on $\sum_{d-2}$ which
satisfies $\nabla^2Y=-QY$, with $\nabla^2$ the Laplace operator on
$\sum_{d-2}$. The eigenvalues for the hyperbolic manifold are
\begin{equation}
Q=\left(\frac{d-3}{2}\right)^2+\xi^2.
\end{equation}
Without any identifications of the pseudosphere the spectrum of
the angular wave equation is continuous, thus $\xi$ takes any real
value $\xi\geq0$. Since the $(d-2)-$dimensional manifold $\sum$ is
a quotient space of the form $H^{d-2}/\Gamma$ and it is a compact
space of constant negative curvature, the spectrum of the angular
wave equation is discretized and thus $\xi$ takes discrete real
values $\xi\geq0$~\cite{Balazs:1986uj}.

 Then the
radial function $R(v)$ becomes
\begin{equation}\label{radial}\
v(1-v)\partial_{v}^2R(v)+\left[1+\left(\frac{d-5}{2}\right)v\right]\partial_{v}R(v)+
\left[\frac{\omega^2}{4pv}-\frac{Q}{4p}-\frac{m^2l^2}{4(1-v)}\right]R(v)=0~.
\end{equation}
Under the decomposition $R(v)=v^\alpha(1-v)^\beta K(v)$, Eq.
(\ref{radial}) can be written as a hypergeometric equation for K
\begin{equation}\label{hypergeometric}\
v(1-v)K''(v)+\left[c-(1+a+b)v\right]K'(v)-ab K(v)=0~.
\end{equation}
Where the coefficients $a$, $b$ and $c$ are given by
\begin{equation}\label{a}\
a=-\left(\frac{d-3}{4}\right)+\alpha+\beta+\frac{i}{2}\sqrt{\frac{\xi^2}{p}+\left(\frac{d-3}{2}\right)^2\left(\frac{1}{p}-1\right)}~,
\end{equation}
\begin{equation}
b=-\left(\frac{d-3}{4}\right)+\alpha+\beta-\frac{i}{2}\sqrt{\frac{\xi^2}{p}+\left(\frac{d-3}{2}\right)^2\left(\frac{1}{p}-1\right)}~,
\end{equation}
\begin{equation}
c=1+2\alpha~,
\end{equation}
where $c$ cannot be an integer and the exponents $\alpha$ and
$\beta$ are
\begin{equation}
\alpha=\pm\frac{i\omega\sqrt{p}}{2p}~,
\end{equation}
\begin{equation}
\beta=\beta_\pm=\left(\frac{d-1}{4}\right)\pm\frac{1}{2}\sqrt{\left(\frac{d-1}{2}\right)^2+m^2l^2}~.
\end{equation}
Without loss of generality, we choose the negative signs for
$\alpha$. The general solution of Eq. (\ref{hypergeometric}) takes
the form
\begin{equation}
K=C_{1}F_{1}(a,b,c;v)+C_2v^{1-c}F_{1}(a-c+1,b-c+1,2-c;v)~,
\end{equation}
which has three regular singular point at $v=0$, $v=1$ and
$v=\infty$. Here, $F_{1}(a,b,c;v)$ is a hypergeometric function
and $C_{1}$, $C_{2}$ are constants. Then, the solution for the
radial function $R(v)$ is
\begin{equation}\label{RV}\
R(v)=C_{1}v^\alpha(1-v)^\beta
F_{1}(a,b,c;v)+C_2v^{-\alpha}(1-v)^\beta
F_{1}(a-c+1,b-c+1,2-c;v)~.
\end{equation}
According to our change of variables at the vicinity of the
horizon when $r\rightarrow r_{+}$, then $z\rightarrow
p-\frac{1}{p}$, $x\rightarrow \frac{1}{p^2}$, $y\rightarrow 1$,
$v\rightarrow 0$ and at the infinity when $r\rightarrow \infty$,
then $z\rightarrow p$, $x\rightarrow 0$, $y\rightarrow 0$,
$v\rightarrow 1$. In the vicinity of the horizon, $v=0$ and using
the property $F(a,b,c,0)=1$, the function $R(v)$ behaves as
\begin{equation}
R(v)=C_1 e^{\alpha \ln v}+C_2 e^{-\alpha \ln v},
\end{equation}
and the scalar field $\varphi$ can be written in the following way
\begin{equation}
\varphi\sim C_1 e^{-i\omega(t+ \frac{\sqrt{p}}{2p}\ln v)}+C_2
e^{-i\omega(t- \frac{\sqrt{p}}{2p}\ln v)}~,
\end{equation}
in which the first term represents an ingoing wave and the second
one an outgoing wave in the black hole. For computing the QNMs, we
have to impose our boundary conditions on the horizon that  there
exist only ingoing waves. This fixes  $C_2=0$. Then the radial
solution becomes
 \begin{equation}\label{horizonsolutiond}
R(v)=C_1 e^{\alpha \ln v}(1-v)^\beta F_{1}(a,b,c;v)=C_1
e^{-i\omega\frac{\sqrt{p}}{2p}\ln v}(1-v)^\beta F_{1}(a,b,c;v)~.
\end{equation}
In order to implement boundary conditions at infinity ($v=1$), we
shall apply in Eq. (\ref{horizonsolutiond}) the Kummer's formula
for the hypergeometric function \cite{M. Abramowitz},
\begin{equation}
F_{1}(a,b,c;v)=\frac{\Gamma(c)\Gamma(c-a-b)}{\Gamma(c-a)\Gamma(c-b)}
F_1(a,b,a+b-c,1-v)+(1-v)^{c-a-b}\frac{\Gamma(c)\Gamma(a+b-c)}{\Gamma(a)\Gamma(b)}F_1(c-a,c-b,c-a-b+1,1-v).
\end{equation}
With this expression the radial function results in
\begin{eqnarray}\label{R}\
\nonumber R(v) &=& C_1 e^{-i\omega\frac{\sqrt{p}}{2p}\ln v}(1-v)^\beta\frac{\Gamma(c)\Gamma(c-a-b)}{\Gamma(c-a)\Gamma(c-b)} F_1(a,b,a+b-c,1-v)\\
&& +C_1 e^{-i\omega\frac{\sqrt{p}}{2p}\ln
v}(1-v)^{c-a-b+\beta}\frac{\Gamma(c)\Gamma(a+b-c)}{\Gamma(a)\Gamma(b)}F_1(c-a,c-b,c-a-b+1,1-v)~.
\end{eqnarray}
The flux is given by
\begin{equation}
\textit{F}=\frac{\sqrt{-g}g^{rr}}{2i}\left(\varphi^{\ast}\partial_{r}\varphi-\varphi\partial_{r}\varphi^{\ast}\right)~,
\label{flux}
\end{equation}
and demanding that it  vanishes at infinity, if
$m^2l^2\geq\frac{25-(d-1)^2}{4}$, results  in a set of two
divergent terms of order $(1-v)^{2\beta-d/2+3}$ and
$(1-v)^{-2\beta+d/2+2}$, for $\beta_-$  and $\beta_+$,
respectively. The condition on the mass of the scalar field agrees
with the Breitenlohner-Freedman condition that any effective mass
must satisfy in order to have a stable AdS asymptotics in
d-dimensions, $m^2l^2\geq-\frac{(d-1)^2}{4}$
\cite{B-F, Mezincescu:1984ev}. It is  worth noting also that for
$d>6$, a negative mass squared for a scalar field is consistent.
Then according to Eq. (\ref{R}), for $\beta_-$, each of these
terms is proportional to
\begin{equation}
\left|\frac{\Gamma(c)\Gamma(c-a-b)}{\Gamma(c-a)\Gamma(c-b)}\right|^2.
\end{equation}
Since the gamma function $\Gamma(x)$ has the poles at $x=-n$ for
$n=0,1,2,...$, the wave function satisfies the considered boundary
condition only upon the following additional restriction
$(c-a)|_{\alpha}=-n$ or $(c-b)|_{\alpha}=-n$ and these conditions
determine the form of the quasi-normal modes as
\begin{equation}\label{qnm}\
\omega=\mp\sqrt{\xi^2+\left(\frac{d-3}{2}\right)^2\left(1-p\right)}-i\sqrt{p}\left(2n+1+\sqrt{\left(\frac{d-1}{2}\right)^2+m^2l^2}\right)~.
\end{equation}
For $\beta_+$, the divergent term is proportional to
\begin{equation}
\left|\frac{\Gamma(c)\Gamma(a+b-c)}{\Gamma(a)\Gamma(b)}\right|^2.
\end{equation}
Since the gamma function $\Gamma(x)$ has the poles at $x=-n$ for
$n=0,1,2,...$, the wave function satisfies the considered boundary
condition only upon the following additional restriction
$(a)|_{\alpha}=-n$ or $(b)|_{\alpha}=-n$ and these conditions
determine the form of the quasi-normal modes as \footnote{
Recently, a preprint appeared in the archives ~\cite{Oliva:2010xn}
 which is
based on an unpublished preprint (CECS-PHY-06/13) by the authors,
in which  the QNMs of a special class of black holes were
presented including the QNMs of Chern-Simons black holes.  Their
method was depending on a suitably rewriting the Chern-Simons
metric as a metric of a massless topological black hole. However,
in our calculation, due to strong coupling problems, we followed
the more physically transparent method of solving the Klein-Gordon
equation in the background of the Chern-Simons black hole. }
\begin{equation}\label{qnm1}\
\omega=\mp\sqrt{\xi^2+\left(\frac{d-3}{2}\right)^2\left(1-p\right)}-i\sqrt{p}\left(2n+1-\sqrt{\left(\frac{d-1}{2}\right)^2+m^2l^2}\right)~.
\end{equation}

We observe that if $p=1$ we recover the QNMs of the massless
topological black holes in $d=4$ dimensions~\cite{Aros:2002te}.
Actually, if $p=1$ then $\mu=0$ and the metric
(\ref{dtopologicalmetric}) coincides with the metric of a massless
topological black hole. It is also interesting to observe that if
$p\neq 1$ the QNMs (\ref{qnm}) and (\ref{qnm1}) of scalar
perturbations of Chern-Simons black holes have the imprint of the
high curvature of the original theory. We expect this to be also
true for the flat and spherical topologies. This result may have
observational implications to the gravitational wave experiments.

According to the AdS/CFT correspondence the relaxation time $\tau$
for a thermal state to reach thermal equilibrium in the boundary
conformal field theory is $\tau=1/\omega_I$ where $\omega_I$ is
the imaginary part of QNMs. As can be seen in relation (\ref{qnm})
$\omega_I$ scales with $\sqrt{p}$. Depending then on the sign of
$\mu$, $p$ can be larger of smaller than one. This means that the
scalar field will decay faster or slower  depending  on the value
of the curvature parameter $k$.

An interesting question is what happens if the scalar field
backreacts on the metric. \footnote{ We thank the referee for
asking this question.} In \cite{Charmousis:2008ce} linear
perturbations were studied of  Gauss-Bonnet maximally symmetric
vacuum solutions. It was found that these solutions  are
perturbatively unstable near the Chern-Simons limit but when this
limit is reached the theory becomes strongly coupled and the
perturbation theory breaks down. If we had considered a coupled
scalar-tensor system this would have resulted in a Chern-Simons
black hole solution with scalar hair. Actually such solutions
exist \cite{Martinez:2004nb} if you think that Chern-Simons black
holes (with $\gamma=-1$) are generalizations of topological black
holes. In a series of papers \cite{papa2} the scalar,
electromagnetic and tensor perturbations were studied of these
hairy black holes. It was found that there exist a critical
temperature below which the vacuum black hole acquires hair. This
happens when the mass $\mu$ goes to zero and the position of the
horizon is at $r_+=1$ (with the AdS length scale $l=1$). Similar
behaviour we expect also for the Chern-Simons black holes. We
speculate that the strong coupling limit is attended when $\mu=0$
which from (\ref{horizon}) is at $r_+=1$. Then it is interesting
to connect this limit with the phase transition observed in
\cite{papa2}. On the other hand notice that in
\cite{Charmousis:2008ce} it was found
 with non-perturbative methods that at the
Chern-Simons limit the Einstein and Gauss-Bonnet vacuum solutions
coexist which is a transient configuration characteristic of a
phase transition. This issue is very interesting and merits
further study, which however is beyond the scope of the present
work.

\section{Mass and area spectrum from quasi-normal modes}

Consider a system with energy $E$ and vibrational frequency
$\omega$
\begin{equation}
\omega=\omega_R=\sqrt{\xi^2+\Big{(}\frac{d-3}{2}\Big{)}^2(1-p)}~.\label{realqnm}
\end{equation}
Define
\begin{equation}
V=-\alpha\left(\frac{d-3}{2}\right)^2\left(2\frac{\Omega_{d-2}}{\Sigma_{d-2}}G_{k}\right)^{\frac{1}{k}},
\end{equation} and then (\ref{realqnm}) becomes
\begin{equation}
\omega=\sqrt{\xi^2+VM^{\frac{1}{k}}}.\label{realnew}
\end{equation}
Assuming that this frequency is a fundamental vibrational
frequency for a black hole of energy $E=M$,  the quantity
\begin{equation}\label{invariant1}\
I_1=\int\frac{dE}{\omega(E)}~,
\end{equation}
is an adiabatic invariant~\cite{Kunstatter:2002pj}. Then Eq.
(\ref{invariant1}), using (\ref{realnew}), has  the following
solution
\begin{equation}\label{invariantHsolution}\
I_1=\frac{M\sqrt{VM^\frac{1}{k}+\xi^2}}{\xi^2}F
_{1}\left[1,\frac{1}{2}+k,1+k,\frac{-VM^{\frac{1}{k}}}{\xi^2}\right].
\end{equation}
Following Hod's proposal \footnote {The Hod's proposal, at least
in the Schwarzschild case, is valid at the large damping limit.
Here, contrary to the Schwarzschild case, the real part of the
QNMs of the Chern-Simons black holes is independent of the mode
number $n$.}, the Bohr-Sommerfeld quantization condition in the
semi-classical limit gives the spectrum
\begin{equation}\label{invariant}\
I_1\approx n\hbar~.
\end{equation}
For the Gauss-Bonnet case with $d=5$ and $k=2$,
 the adiabatic invariant Eq.
(\ref{invariantHsolution}) can be written as
\begin{equation}
I_1=\frac{4\left(B^2M-B\sqrt{M}\xi^2+2\left(-1+\sqrt{1+\frac{B\sqrt{M}}{\xi^2}}\right)\xi^4\right)}{3B^2\sqrt{1+B\sqrt{M}}}~,
\end{equation}
where
\begin{equation}
B=-\alpha\left(2\frac{\Omega_{3}}{\Sigma_{3}}G_{2}\right)^{\frac{1}{2}}.
\end{equation}
To simplify the above expression, without loosing the generality,
we choose $\xi=0$. Then equating the above equation with Eq. (\ref{invariant}) we obtain the mass spectrum
\begin{equation}
M(n)=\frac{1}{4}\Big{(}3n\hbar\sqrt{\frac{B}{2}}\Big{)}^\frac{4}{3}~.\label{massspectrum}
\end{equation}
It is worth noting that for $k=2$, $\alpha$ can take the values
$\pm1$. To find the area spectrum, we use the horizon area of the
black hole, that is given by
\begin{equation}\label{horizonarea}\
A_{r_+}=\Sigma_3r_{+}^3~,
\end{equation}
where $r_{+}$ is given by Eq. (\ref{horizon}). Then using the mass
spectrum (\ref{massspectrum}), the area spectrum becomes
\begin{equation}
A_n=\Sigma_3
l^3\Big{(}1-\frac{B}{2}\Big{(}3n\hbar\sqrt{\frac{B}{2}}\Big{)}^\frac{2}{3}\Big{)}^\frac{3}{2}~.
\label{qarea}
\end{equation}
We observe that the mass and area spectrum given by
(\ref{massspectrum}) and (\ref{qarea}) respectively, are not
evenly spaced. The no equidistance of the spectrum was also found
in other black hole cases. The (2+1)-dimensional BTZ black hole
was studied in connection to the Hod's
conjecture~\cite{Birmingham:2003wa}. It was found that there is a
connection between the quasi-normal modes and the quantum (2+1)
black holes but it was not found a quantization of the horizon
area. Also in~\cite{Setare:2003hm} a quantization of the horizon
area was found for a non-rotating BTZ black hole but it is not
evenly spaced. In~\cite{Kothawala:2008in} the high curvature
Lanczos-Lovelock theories were studied and it was found that
 the area spacing is
not equidistant and it was claimed that the notion of quantum of
entropy is more natural in these theories. Also, in acoustic
(2+1)-dimensional black holes, the mass and area spectrum  is not
evenly spaced~\cite{Lepe:2004kv}.

It is interesting to note that the complexity of the horizon of
the hyperbolic geometry has an important effect on the mass and
area spectrum of the Chern-Simons black holes. The $B$ factor that
appears in (\ref{massspectrum}) and (\ref{qarea}) is inverse
proportional to the volume of the hyperbolic space $\Sigma_{3}$.
For high genus the volume $\Sigma_{3}$ can be arbitrary
large~\cite{HyperB} giving an irregular spacing,  leading
eventually to a breakdown of the quantization conditions
(\ref{massspectrum}) and (\ref{qarea}).

The same behaviour also appears in the Maggiore's approach
\cite{Maggiore:2007nq}. In this approach we identify $\omega
\simeq \omega_I$ of QNMs in the high damping limit (large $n$
limit) and evaluate the adiabatic expression
\begin{equation}\label{invariantM1}\
I_2=\int\frac{dM}{\omega_t}~,
\end{equation}
where, the transition frequency $\omega_t$ is given by
\begin{equation}\label{transitionfrequency}\
\omega_t=|(\omega_I)_n|-|(\omega_I)_{n-1}|=2\sqrt{p}=2\sqrt{1+CM^\frac{1}{k}},
\end{equation}
where
\begin{equation}
C=\alpha\left(2\frac{\Omega_{d-2}}{\Sigma_{d-2}}G_{k}\right)^{\frac{1}{k}},
\end{equation}
and
\begin{equation}
\omega_I(n\rightarrow\infty)=-2n\sqrt{p}~.
\end{equation}
Thus, Eq.~(\ref{invariantM1}) has the following solution
\begin{equation}\label{invariantMsolution}\
I_2=\frac{M}{2}F
_{2}\left[\frac{1}{2},k,1+k,-CM^{\frac{1}{k}}\right].
\end{equation}
In the specific example of $d=5$ and $k=2$, imposing the
Bohr-Sommerfeld quantization condition the adiabatic invariant Eq.
(\ref{invariantMsolution}) becomes
\begin{equation}
I_2=\frac{2\left[2-\sqrt{1-B\sqrt{M}}\left(B\sqrt{M}+2\right)\right]}{3B^2}~,\label{magsol}
\end{equation}
where $B=-C$. The solution (\ref{magsol}) gives a complicated mass
and area spectrum that they are not evenly spaced and they depend
also on the factor $B$.

Having evaluated the mass and area spectrum we can ask if this
information can be used to evaluate the spectrum of the entropy of
Chern-Simons black holes. In Einstein's theory this is possible
using the Bekenstein's entropy formula. However, in Lovelock
theories the entropy is not proportional to the horizon area. The
reason is that the entropy  has a leading Lovelock correction term
which appears as an induced curvature term which is nothing else
but the Euler density of the horizon surface \cite{Clunan:2004tb}.
In light of this, can the entropy spectrum of  Chern-Simons black
holes be evenly spaced in spite that their area spectrum is not?
In \cite{Kothawala:2008in} a general argument is given that in
Lovelock theories  the entropy should be quantized with an equally
spaced spectrum and they demonstrated this argument in the
five-dimensional Gauss-Bonnet theory using the asymptotic form of
the QNMs. We believe that the effect of Lovelock gravity on the
entropy spectrum is a pure geometrical effect and considering that
the Chern-Simons black holes depend explicitly on the curvature
parameter $k$,   a careful analysis should be carried out in the
lines of \cite{Clunan:2004tb} to settle this issue.

\section{Reflection, transmission coefficients and absorption cross sections}

The reflection and the transmission coefficients are defined by
\begin{equation}\label{reflectiond}\
\Re :=\left|\frac{F_{\mbox{\tiny asymp}}^{\mbox{\tiny
out}}}{F_{\mbox{\tiny asymp}}^{\mbox{\tiny in}}}\right|, \qquad
\mbox{and} \qquad  \mathfrak{U}:=\left|\frac{F_{\mbox{\tiny
hor}}^{\mbox{\tiny in}}}{F_{\mbox{\tiny asymp}}^{\mbox{\tiny
in}}}\right|,
\end{equation}
where $F$ is the flux given in Eq. (\ref{flux}) adapted for the
radial function $R(v)$
\begin{equation}\label{fluxd}\
\textit{F}=\frac{\sqrt{-g}g^{rr}}{2i}\left(R^{\ast}\partial_{r}R-R\partial_{r}R^{\ast}\right).
\end{equation}

To calculate the above coefficients we need to know the behaviour
of the radial function both at the horizon and at the asymptotic
infinity. The behaviour at the horizon is given by Eq. (\ref{horizonsolutiond}) and using Eq. (\ref{fluxd}), we get the
flux at the horizon up to an irrelevant factor  from the angular
part of the solution
\begin{equation}
\textit{F}
_{hor}^{in}=-\left|C_{1}\right|^{2}\omega l^{d-3}p^{\frac{d-4}{2}}~.
\end{equation}

To obtain the asymptotic behaviour of the $R(v)$ we use
$1-v=\frac{l^2p}{r^2}$ and taking into account the limit of
$R(v)$, Eq. (\ref{RV}), when $v\rightarrow1$, we have
\begin{equation}\label{Rinfinity}\
R(r) =
C_1\left(\frac{l\sqrt{p}}{r}\right)^{2\beta}\frac{\Gamma(c)\Gamma(c-a-b)}{\Gamma(c-a)\Gamma(c-b)}
+C_1\left(\frac{l\sqrt{p}}{r}\right)^{d-1-2\beta}\frac{\Gamma(c)\Gamma(a+b-c)}{\Gamma(a)\Gamma(b)}~.
\end{equation}
On the other hand, when $r\rightarrow\infty$, Eq. (\ref{waveequation}) approximates to
\begin{equation}\label{radiald}\
\partial_ {r}^2R(r)+\frac{d}{r}\partial_ {r}R(r)+\left(\frac{\omega^2l^2}{r^4}-\frac{Ql^2}{r^4}-\frac{m^2l^2}{r^2}\right)R(r)=0~,
\end{equation}
where, we have used the ansatz $\varphi=R(r)Y(\sum)e^{-i\omega
t}$. The behaviour at the asymptotic region is the same as for the
Topological massless black holes \cite{{Gonzalez:2010ht}}. The
solution of Eq. (\ref{radiald}) is a linear combination of the
Bessel function \cite{M. Abramowitz}\ given by
\begin{equation}
R(r)=\left(\frac{\sqrt{A}}{2r}\right)^{\frac{d-1}{2}}\left[D_1\Gamma{(1-C)}J_{-C}\left(\frac{\sqrt{A}}{r}\right)+D_2\Gamma{(1+C)}J_{C}\left(\frac{\sqrt{A}}{r}\right)\right],
\end{equation}
where
\begin{equation}
A=l^2(l^2\omega^2-Q)~,
\end{equation}
\begin{equation}
C=\frac{1}{2}\sqrt{(d-1)^2+4m^2l^2}~.
\end{equation}
Now, using the expansion of the Bessel function \cite{M.
Abramowitz}\
\begin{equation}
J_n(x)=\frac{x^n}{2^n\Gamma{(n+1)}}\left\{1-\frac{x^2}{2(2n+2)}+...\right\},
\end{equation}
for $x\ll1$, we find the asymptotic solution in the polynomial
form
\begin{equation}\label{Rasympd}\
R_{asymp.}(r)=D_1\left(\frac{\sqrt{A}}{2r}\right)^{\frac{d-1}{2}-C}+D_2\left(\frac{\sqrt{A}}{2r}\right)^{\frac{d-1}{2}+C},
\end{equation}
for $\frac{\sqrt{A}}{r}\ll1$. Introducing,
\begin{equation}
\widehat{D}_1\equiv
D_1\left(\frac{\sqrt{A}}{2}\right)^{\frac{d-1}{2}-C},\,\,\,
\widehat{D}_2\equiv
D_2\left(\frac{\sqrt{A}}{2}\right)^{\frac{d-1}{2}+C}~,
\end{equation}
we write Eq. (\ref{Rasympd}) as
\begin{equation}\label{Rasympd1}\
R_{asymp.}(r)=\widehat{D}_1\left(\frac{1}{r}\right)^{\frac{d-1}{2}-C}+\widehat{D}_2\left(\frac{1}{r}\right)^{\frac{d-1}{2}+C}.
\end{equation}

In Ref.~\cite{Hertog:2004rz} it was discussed that a scalar field
with asymptotic behaviour similar to that of Eq. (\ref{Rasympd1}),
generically may lead to an unstable state (the so-called, big
crunch singularity) which is a clear indication of the appearance
of a nonlinear instability. However,  such  an instability
influences  the boundary conditions that scalar fields must
satisfy at infinity~\cite{Mezincescu:1984ev}. Although the
modified boundary conditions preserve the full set of asymptotic
AdS symmetries and allow for a finite conserved energy to be
defined, this energy can be negative. We notice, that the
imposition of regularity condition on the radial function
(\ref{Rasympd1}) at the infinity implies $\frac{d-1}{2}-C\geq0$ or
$-\frac{(d-1)^2}{4}\leq m^2l^2\leq 0$. This is in agreement with
the condition for the mass in order to have a stable asymptotic
AdS spacetime in $d$-dimensions, $m^2l^2\geq-\frac{(d-1)^2}{4}$
\cite{Mezincescu:1984ev}. Besides, $a+b-c=-C$, for
$\beta=\beta_-$, and $c-a-b=-C$, for $\beta=\beta_+$. For this
reason $C$ can not be an integer, because the gamma function is
singular at that point and the regularity conditions are not
satisfied.

 We now take advantage of the inherent symmetry that the
radial solution possesses in the asymptotic region.  More
specifically, we have the freedom to choose the form of the
constant $\beta$ since by changing  $\beta_+$ to $\beta_-$  this
solution is unchanged. Comparison of Eqs. (\ref{Rinfinity}) and
(\ref{Rasympd1}), regarding $\beta = \beta_-$,  allows us to
immediately  read off the coefficients $\widehat{D}_1$ and
$\widehat{D}_2$
\begin{eqnarray}\label{Dbarra1}\
\widehat{D}_1&=&C_1 \left(l\sqrt{p}\right)^{2\beta_-}\frac{\Gamma(c)\Gamma(c-a-b)}{\Gamma(c-a)\Gamma(c-b)}~,\\
\label{Dbarra2}\ \widehat{D}_2&=& C_1
\left(l\sqrt{p}\right)^{d-1-2\beta_-}\frac{\Gamma(c)\Gamma(a+b-c)}{\Gamma(a)\Gamma(b)}~.
\end{eqnarray}

Therefore, the behaviour at the asymptotic region is given by Eq. (\ref{Rasympd1}) and using Eq. (\ref{fluxd}), we get the flux up
to an irrelevant factor from the angular part of the solution and
it is given by
\begin{equation}\label{fluxdinfinity}\
\textit{F}
_{asymp.}=-iC\left(\frac{1}{l^2}-\frac{p}{r^2}\right)\left(\widehat{D}_{2}^*\widehat{D}_{1}-\widehat{D}_{1}^*\widehat{D}_{2}\right).
\end{equation}

We notice here that  the distinction between the ingoing and
outgoing fluxes at the asymptotic region is a non trivial task
because the spacetime is asymptotically AdS. In order to
characterize the fluxes is convenient to split up the coefficients
$\widehat{D}_1$ and $\widehat{D}_2$ in terms of the incoming and
outgoing coefficients, $D_{\mbox{\tiny in}}$ and $D_{\mbox{\tiny
out}}$, respectively. We define  $\widehat{D}_1 = D_{\mbox{\tiny
in}} + D_{\mbox{\tiny out}}$ and $\widehat{D}_2 = i h
(D_{\mbox{\tiny out}} - D_{\mbox{\tiny in}})$ with $h$ being a
dimensionless constant which will be assumed to be independent of
the energy $\omega$ \cite{Birmingham:1997rj,Kim:1999un, Oh:2008tc,
Kao:2009fh}.  In this way, we rewrite the asymptotic flux Eq.
(\ref{fluxdinfinity}) as

\begin{equation}
\textit{F} _{asymp.}\approx
\frac{2hC}{l^2}\left(\left|D_{\mbox{\tiny
in}}\right|^{2}-\left|D_{\mbox{\tiny out}}\right|^{2}\right)~.
\end{equation}
Therefore, the reflection and transmission coefficients
 are given by
\begin{equation}
\Re=\frac{\left|D_{\mbox{\tiny
out}}\right|^2}{\left|D_{\mbox{\tiny in}}\right|^2}~,\label{coef1}
\end{equation}
\begin{equation}
\mathfrak{U}=\frac{\omega
l^{d-1}p^{\frac{d-4}{2}}\left|C_{1}\right|^2}{2\left|h\right|C\left|D_{\mbox{\tiny
in}}\right|^2}~,\label{coef2}
\end{equation}
and the absorption cross section, $\sigma_{abs}$, is given by
\begin{equation}\label{absorptioncrosssection}\
\sigma_{abs}=\frac{\mathfrak{U}}{\omega}=\frac{l^{d-1}p^{\frac{d-4}{2}}\left|C_{1}\right|^2}{2\left|h\right|C\left|D_{\mbox{\tiny
in}}\right|^2}~,
\end{equation}
where, the coefficients $D_{\mbox{\tiny in}}$ and $D_{\mbox{\tiny
out}}$ are given by
\begin{equation}\label{D11}\
 D_{\mbox{\tiny in}}= \frac{C_1}{2}\left[\left(l\sqrt{p}\right)^{2\beta_-}\frac{\Gamma(c)\Gamma(c-a-b)}{\Gamma(c-a)\Gamma(c-b)}+\frac{i}{h}\left(l\sqrt{p}\right)^{d-1-2\beta_-}\frac{\Gamma(c)\Gamma(a+b-c)}{\Gamma(a)\Gamma(b)}\right],
 \end{equation}
\begin{equation}\label{D21}\
 D_{\mbox{\tiny out}}=
 \frac{C_1}{2}\left[\left(l\sqrt{p}\right)^{2\beta_-}\frac{\Gamma(c)\Gamma(c-a-b)}{\Gamma(c-a)\Gamma(c-b)}-\frac{i}{h}\left(l\sqrt{p}\right)^{d-1-2\beta_-}\frac{\Gamma(c)\Gamma(a+b-c)}{\Gamma(a)\Gamma(b)}\right].
 \end{equation}

We will carry out a  numerical analysis of the reflection and
transmission coefficients of Eq. (\ref{coef1}) and Eq. (\ref{coef2}) for a five-dimensional Chern-Simons black hole.
 For our numerics we will chose  $p=0.5,2$, which indicates a black
hole with $\alpha=\mp1$, respectively. It is worth to note that
this choice corresponds to  different masses for the black holes.
In the literature, there are various ways to chose the constant
$h$. It can be chosen  so that the absorption cross section can be
expressed by the area of horizon in the
 zero-frequency limit \cite{Birmingham:1997rj, Das:1996we}.  Also, it can be chosen to obtain the usual value
  of the Hawking temperature \cite {Kim:1999un}
 or in such a way so that the sum of the reflection coefficient and the transmission coefficient be unity \cite {Kim:2004sf}.

\begin{figure}
\includegraphics[width=4.0in,angle=0,clip=true]{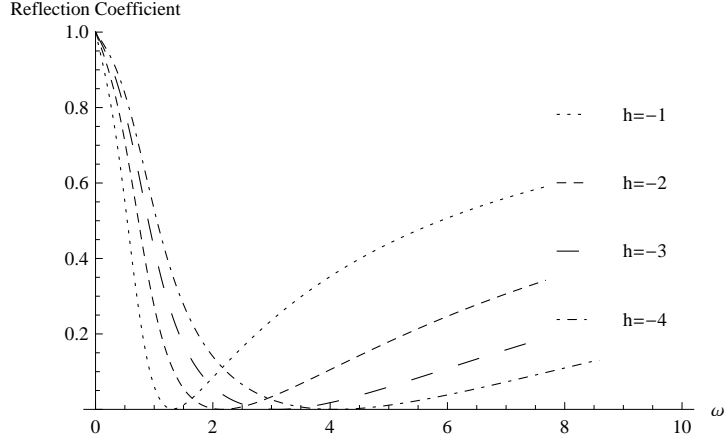}
\caption{Reflection coefficient v/s $\omega$; $d=5$,
$m^2l^2=-15/4$, $l=1$, $p=0.5$ and $\xi=0$.}
\label{ReflectionCoefficientCSBH5dhA}
\end{figure}

\begin{figure}
\includegraphics[width=4.0in,angle=0,clip=true]{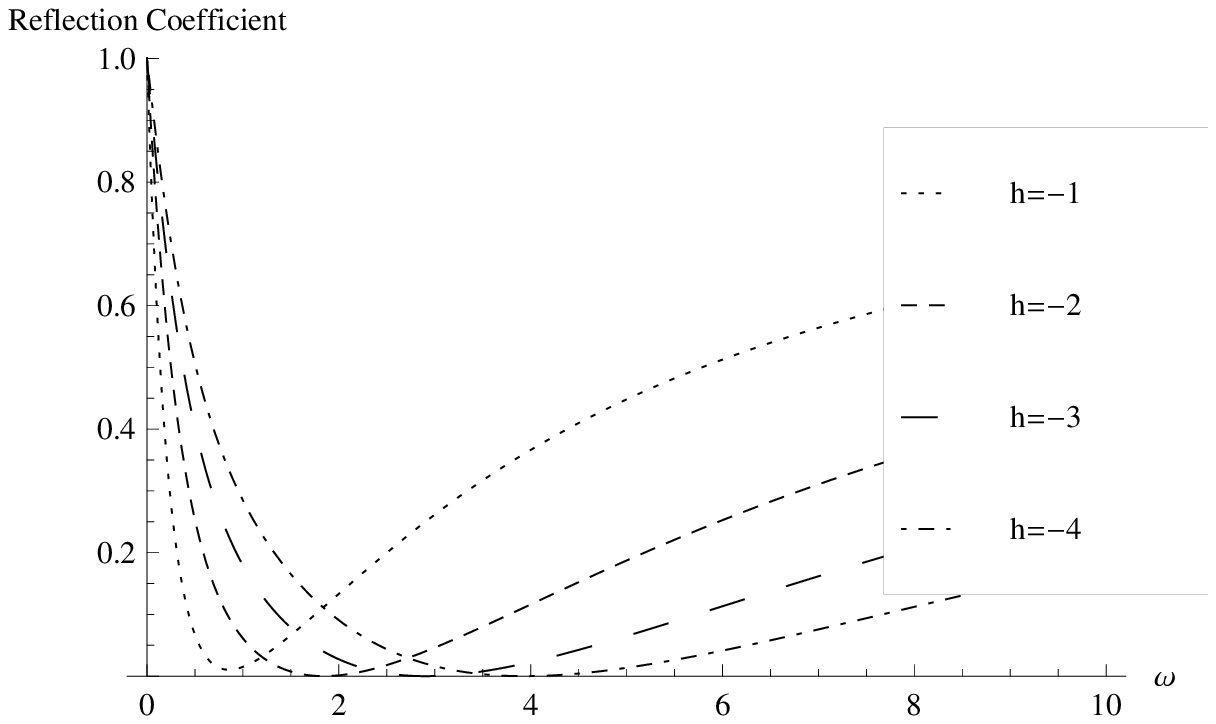}
\caption{Reflection coefficient v/s $\omega$; $d=5$,
$m^2l^2=-15/4$, $l=1$, $p=2$ and $\xi=0$.}
\label{ReflectionCoefficientCSBH5dhB}
\end{figure}

\begin{figure}
\includegraphics[width=4.0in,angle=0,clip=true]{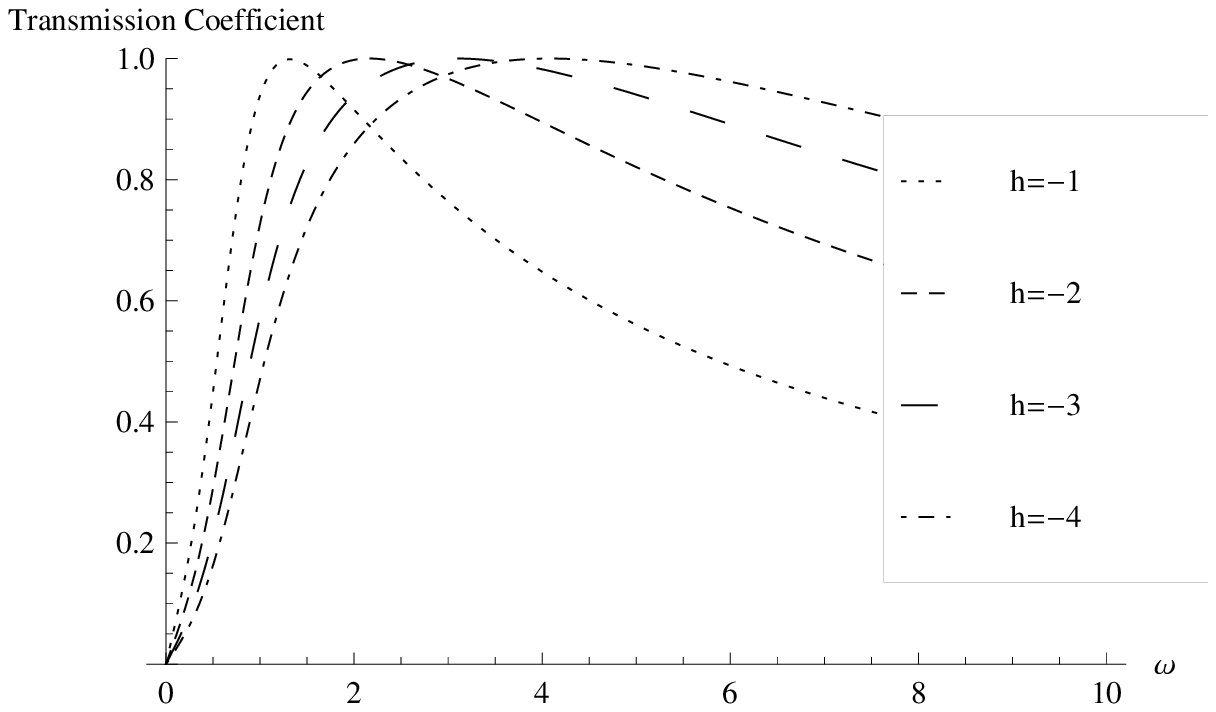}
\caption{Transmission coefficient v/s $\omega$; $d=5$,
$m^2l^2=-15/4$, $l=1$, $p=0.5$ and $\xi=0$.}
\label{TransmissionCoefficientCSBH5dhA}
\end{figure}

\begin{figure}
\includegraphics[width=4.0in,angle=0,clip=true]{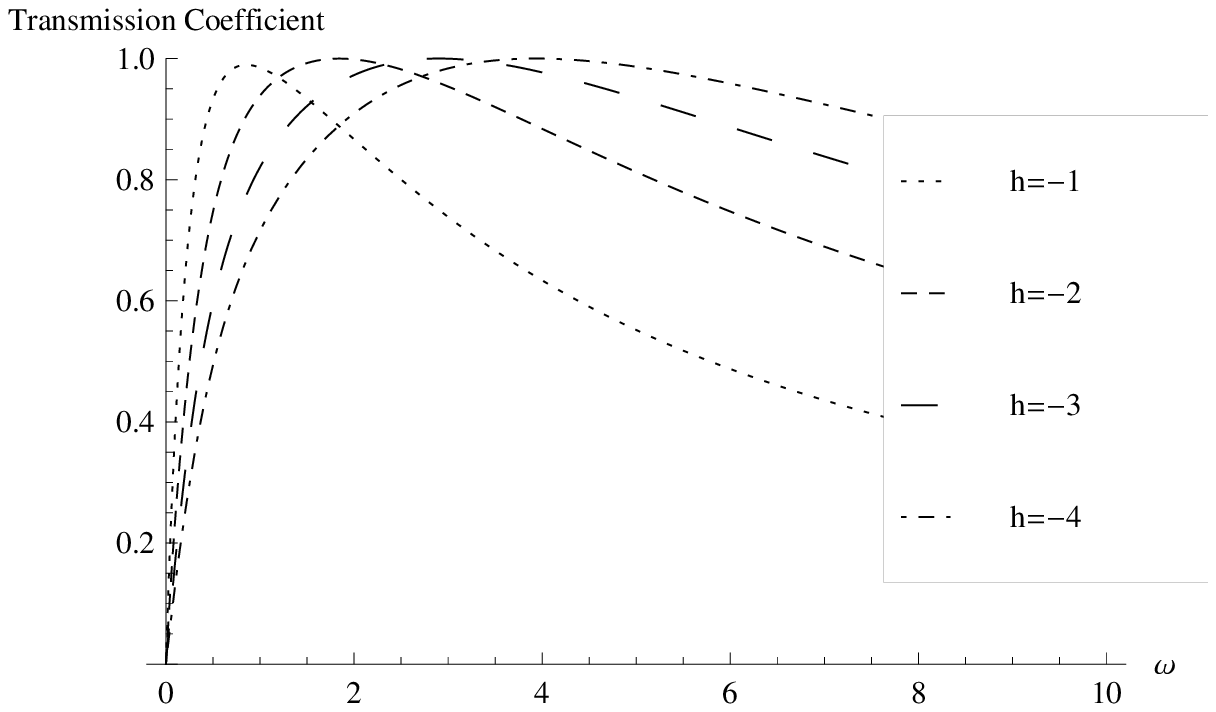}
\caption{Transmission coefficient v/s $\omega$; $d=5$,
$m^2l^2=-15/4$, $l=1$, $p=2$ and $\xi=0$.}
\label{TransmissionCoefficientCSBH5dhB}
\end{figure}

\begin{figure}
\includegraphics[width=4.0in,angle=0,clip=true]{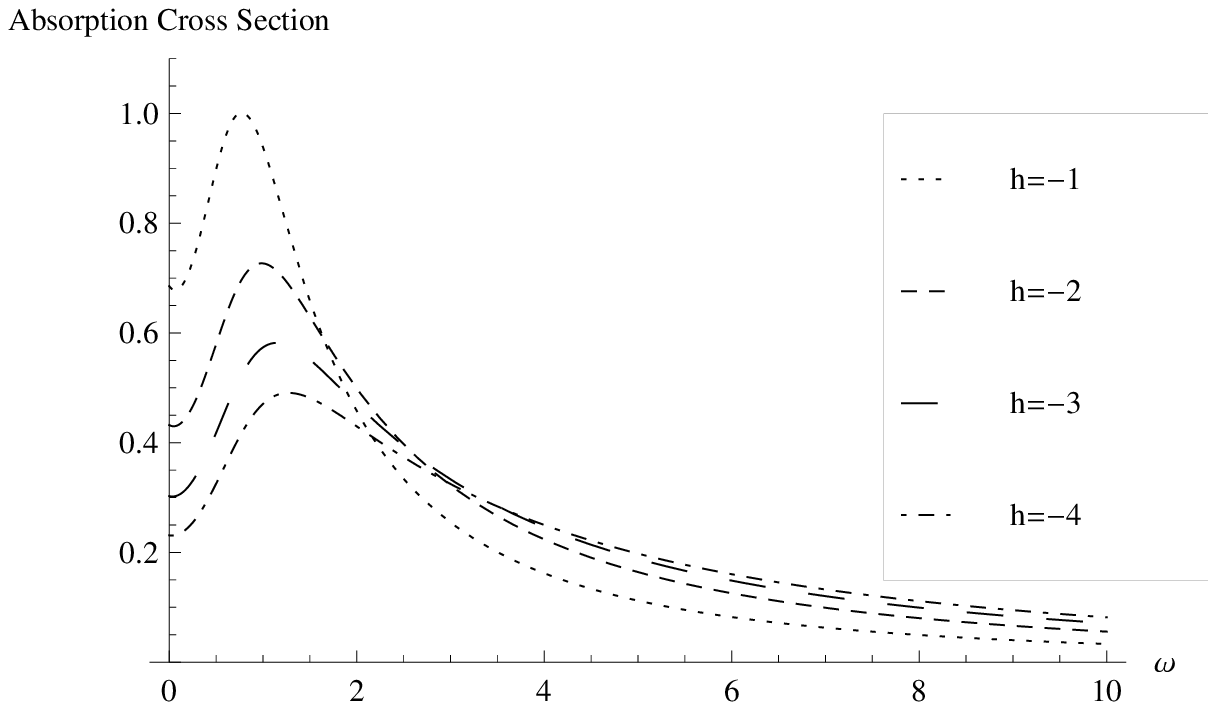}
\caption{Absorption Cross Section v/s $\omega$; $d=5$,
$m^2l^2=-15/4$, $l=1$, $p=0.5$ and $\xi=0$.}
\label{AbsorptionCrossSectionCSBH5dhA}
\end{figure}

\begin{figure}
\includegraphics[width=4.0in,angle=0,clip=true]{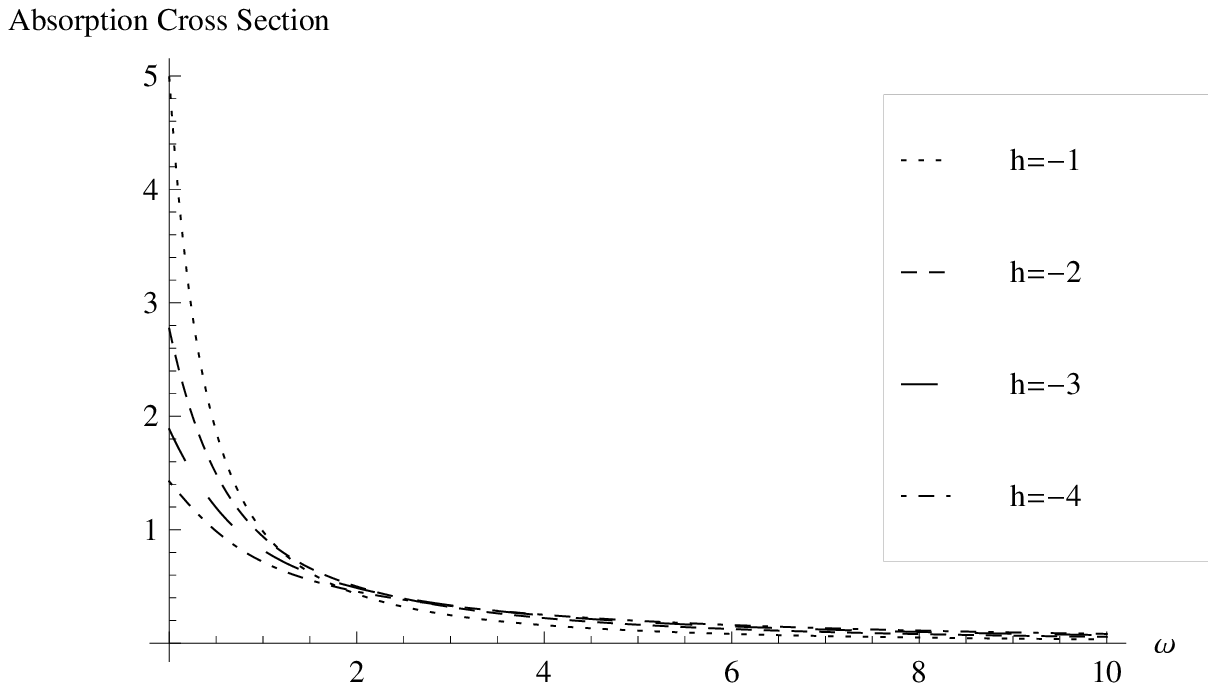}
\caption{Absorption Cross Section v/s $\omega$; $d=5$,
$m^2l^2=-15/4$, $l=1$, $p=2$ and $\xi=0$.}
\label{AbsorptionCrossSectionCSBH5dhB}
\end{figure}

In view of these uncertainties, we leave $h$ as a free parameter
and our only requirement is that the reflection and transmission
coefficients to be self-consistent so that the calculated greybody
factors
 to have a physical meaning \cite{Oh:2008tc}.
  In consequence, for  different values of $h$ and for $m^2l^2=-15/4$, $l=1$, $p=0.5,2$ and
  $\xi=0$,
  we plot the reflection
 coefficient Fig.~(\ref{ReflectionCoefficientCSBH5dhA}, \ref{ReflectionCoefficientCSBH5dhB}), the transmission coefficient Fig.~(\ref{TransmissionCoefficientCSBH5dhA}, \ref{TransmissionCoefficientCSBH5dhB})
  and the greybody
 factors, Fig.~(\ref{AbsorptionCrossSectionCSBH5dhA}, \ref{AbsorptionCrossSectionCSBH5dhB}).
   Essentially, we found the same general
 behaviour for the different values of $h$ and the only difference is a shift in the location of the minimum or maximum of the reflection
 and transmission coefficients respectively.
 Also we observe  that the parameter $h$ must be less than zero and greater than some value such that the
  absorption cross section or the greybody factor
 be real in the zero-frequency limit, otherwise it is imaginary. However in the zero-frequency
 limit,
  the greybody factors depend on the value
 of $h$, so that the coefficient is increasing if the parameter $h$ is increasing, as it can be seen in
 Fig.~(\ref{AbsorptionCrossSectionCSBH5dhA},
 \ref{AbsorptionCrossSectionCSBH5dhB}).
 If we plot the combination
  $\Re+\mathfrak{U}$ for $m^2l^2=-15/4$, $l=1$, $p=0.5,2$,
 $\xi=0$ and $h=-1,-2,-3,-4$ using the
 Fig.~(\ref{ReflectionCoefficientCSBH5dhA}, \ref{TransmissionCoefficientCSBH5dhA})
 and Fig.~(\ref{ReflectionCoefficientCSBH5dhB},
 \ref{TransmissionCoefficientCSBH5dhB}) we get the  $\Re+\mathfrak{U}=1$
 in accordance with
  \cite {Kim:2004sf}. As we discussed previously,  our choice of $m^2l^2$  is in agreement with
the condition for the mass in order to have a stable asymptotic
AdS spacetime in five dimensions.

\begin{figure}
\includegraphics[width=4.0in,angle=0,clip=true]{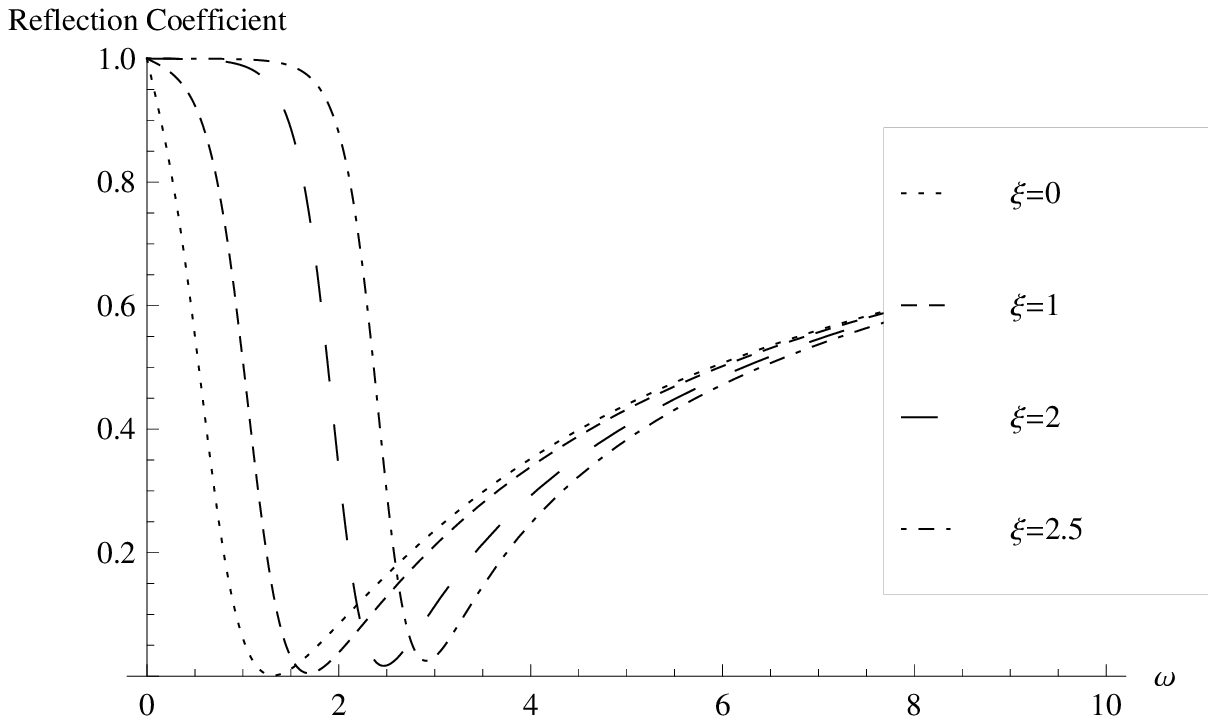}
\caption{Reflection coefficient v/s $\omega$; $d=5$,
$m^2l^2=-15/4$, $l=1$, $p=0.5$ and $h=-1$.}
\label{ReflectionCoefficientCSBH5dA}
\end{figure}

\begin{figure}
\includegraphics[width=4.0in,angle=0,clip=true]{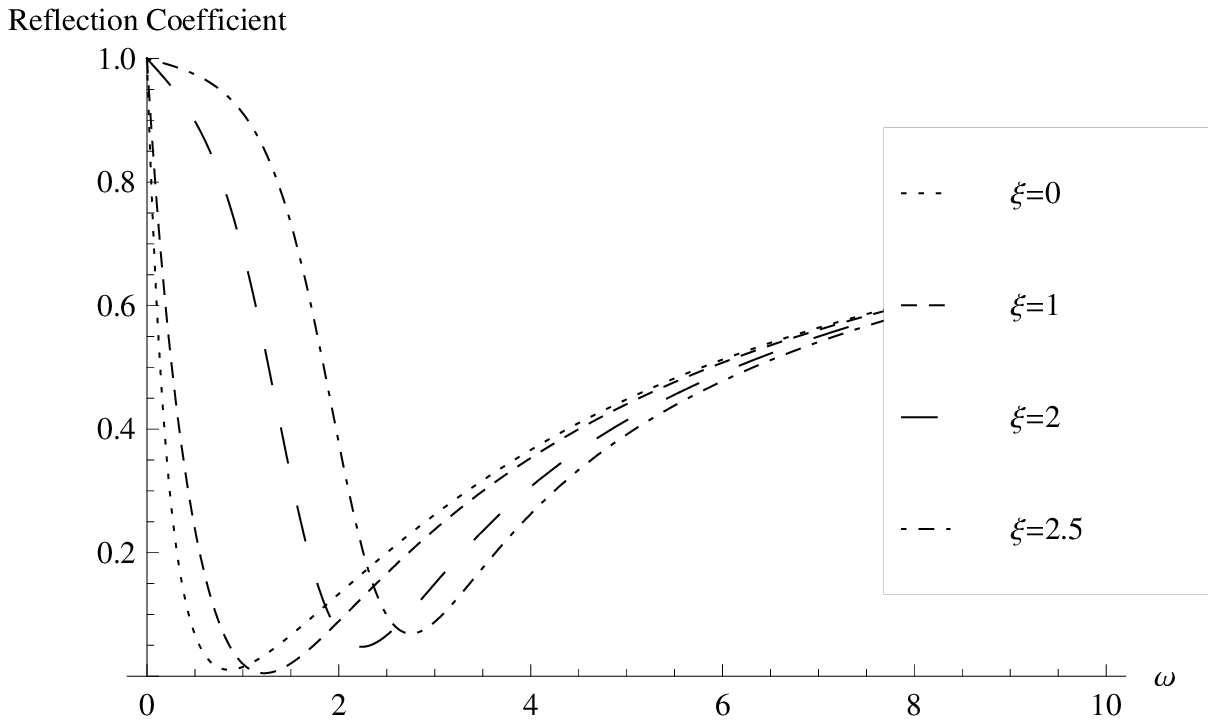}
\caption{Reflection coefficient v/s $\omega$; $d=5$,
$m^2l^2=-15/4$, $l=1$, $p=2$ and $h=-1$.}
\label{ReflectionCoefficientCSBH5dB}
\end{figure}

\begin{figure}
\includegraphics[width=4.0in,angle=0,clip=true]{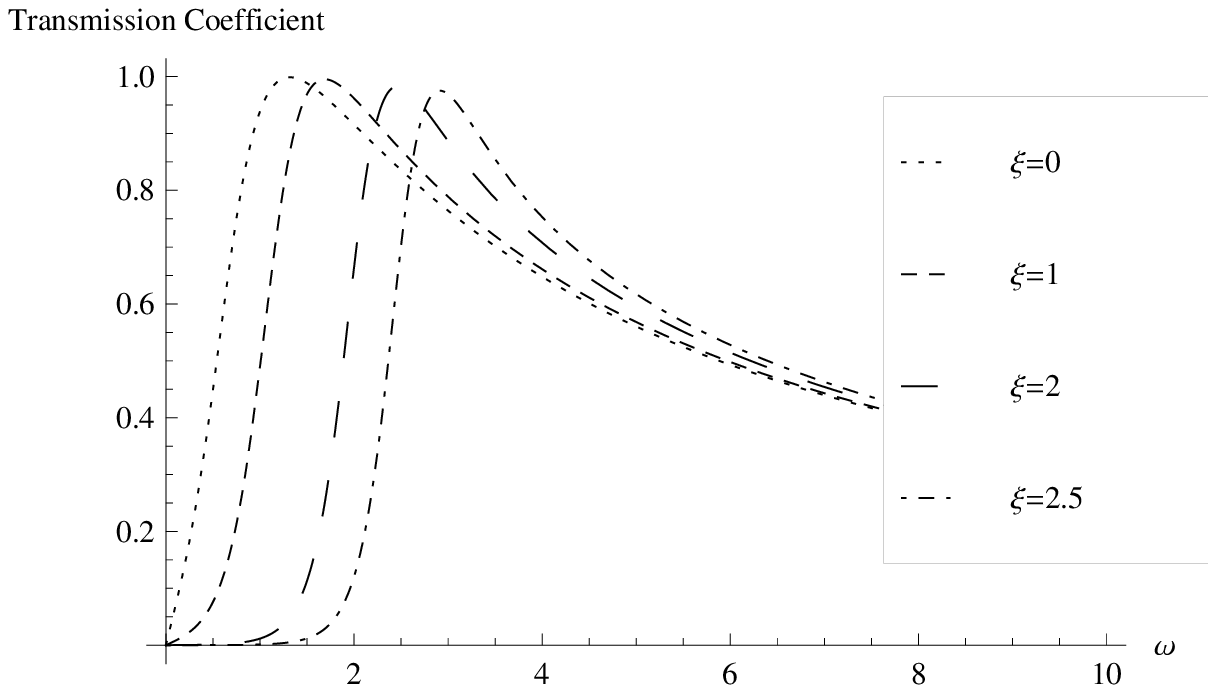}
\caption{Transmission coefficient v/s $\omega$; $d=5$,
$m^2l^2=-15/4$, $l=1$, $p=0.5$ and $h=-1$.}
\label{TransmissionCoefficientCSBH5dA}
\end{figure}

\begin{figure}
\includegraphics[width=4.0in,angle=0,clip=true]{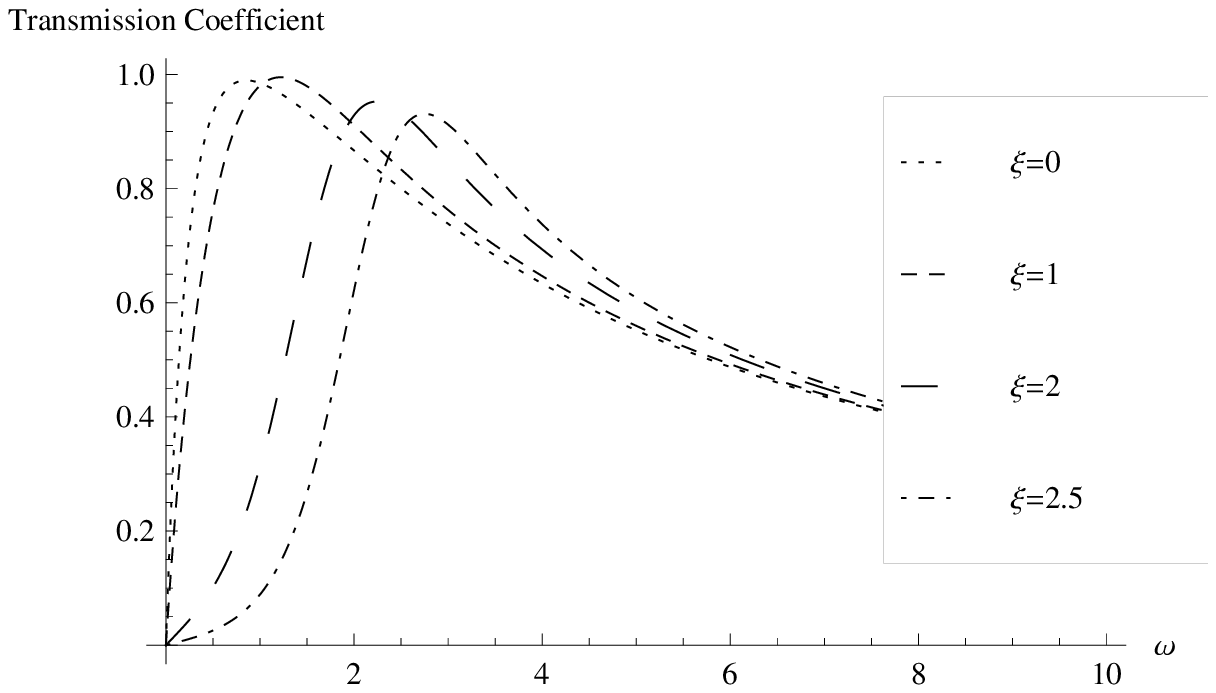}
\caption{Transmission coefficient v/s $\omega$; $d=5$,
$m^2l^2=-15/4$, $l=1$, $p=2$ and $h=-1$.}
\label{TransmissionCoefficientCSBH5dB}
\end{figure}

\begin{figure}
\includegraphics[width=4.0in,angle=0,clip=true]{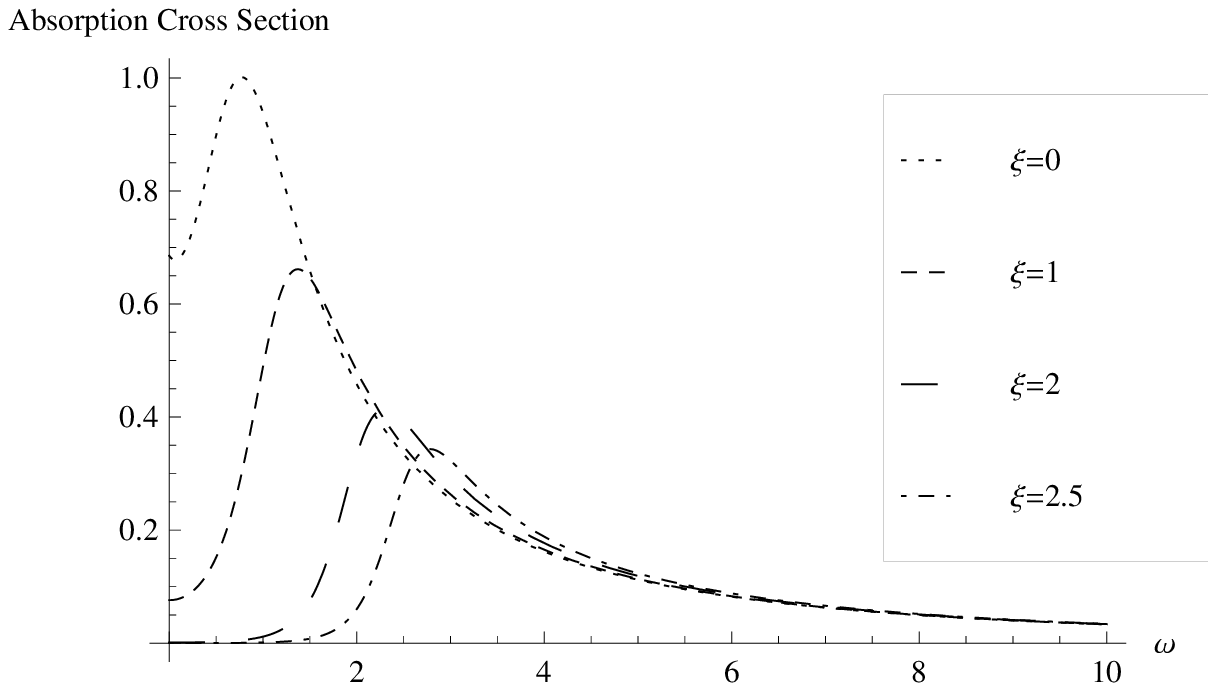}
\caption{Absorption Cross Section v/s $\omega$; $d=5$,
$m^2l^2=-15/4$, $l=1$, $p=0.5$ and $h=-1$.}
\label{AbsorptionCrossSectionCSBH5dA}
\end{figure}

\begin{figure}
\includegraphics[width=4.0in,angle=0,clip=true]{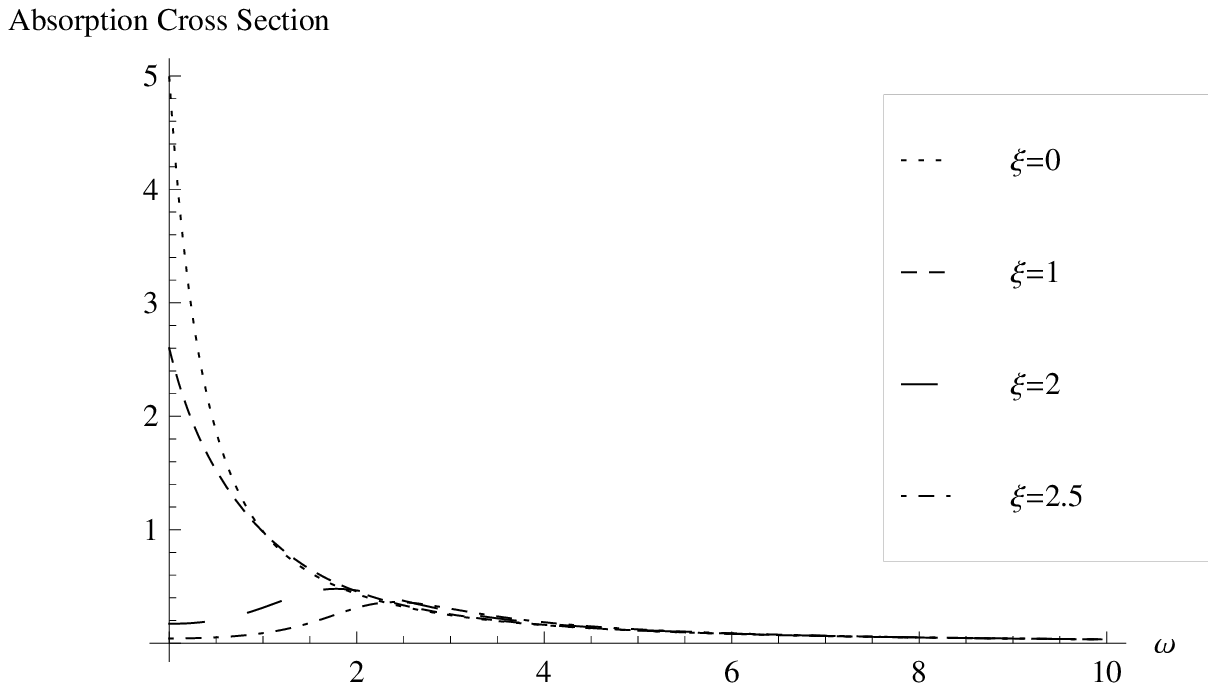}
\caption{Absorption Cross Section v/s $\omega$; $d=5$,
$m^2l^2=-15/4$, $l=1$, $p=2$ and $h=-1$.}
\label{AbsorptionCrossSectionCSBH5dB}
\end{figure}

\begin{figure}
\includegraphics[width=4.0in,angle=0,clip=true]{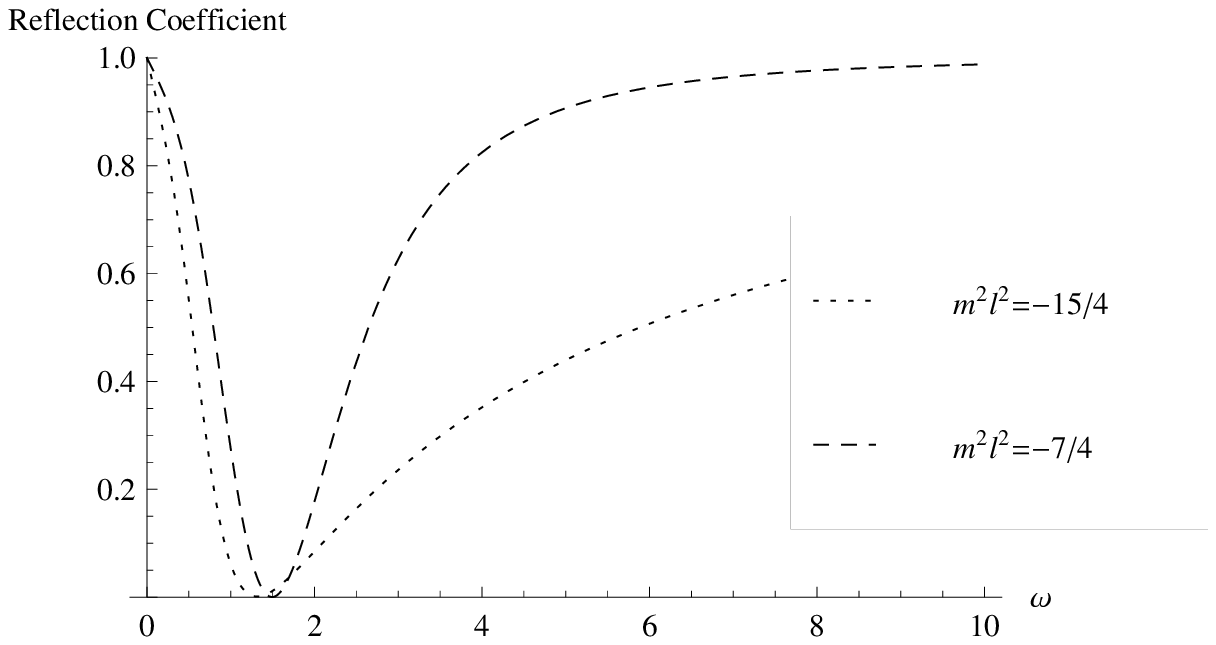}
\caption{Reflection coefficient v/s $\omega$; $d=5$,
$m^2l^2=-15/4,-7/4$, $l=1$, $h=-1$, $p=0.5$ and $\xi=0$.}
\label{ReflectionCoefficientCSBH5dmA}
\end{figure}

\begin{figure}
\includegraphics[width=4.0in,angle=0,clip=true]{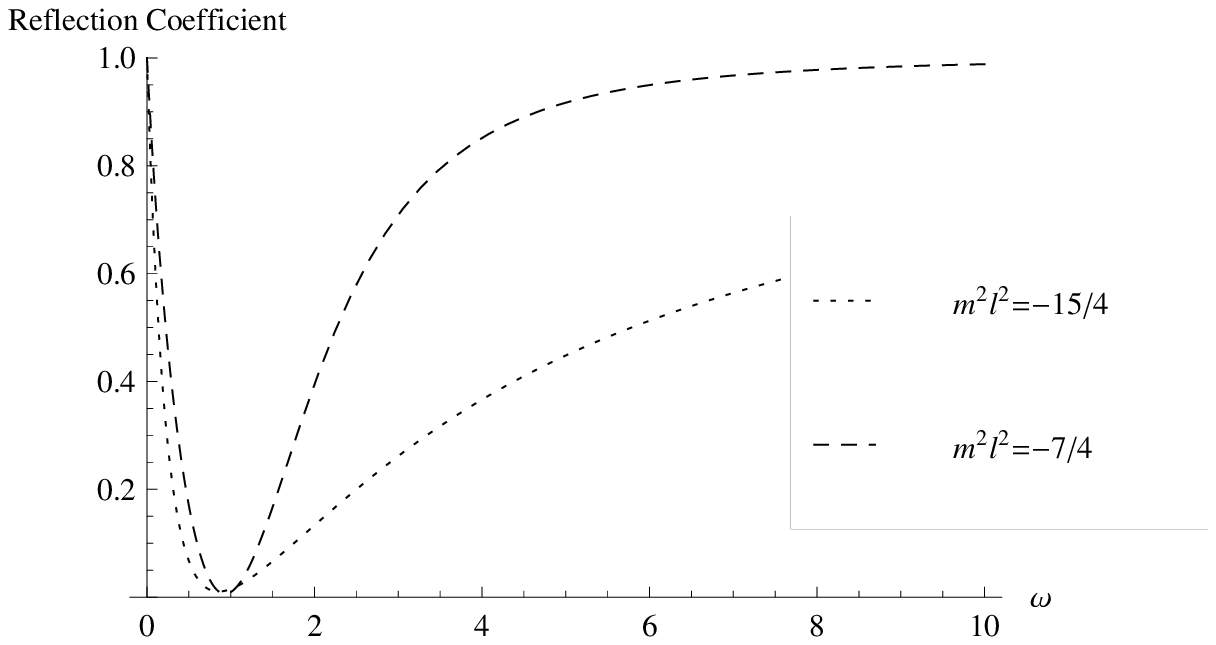}
\caption{Reflection coefficient v/s $\omega$; $d=5$,
$m^2l^2=-15/4,-7/4$, $l=1$, $h=-1$, $p=2$ and $\xi=0$.}
\label{ReflectionCoefficientCSBH5dmB}
\end{figure}

\begin{figure}
\includegraphics[width=4.0in,angle=0,clip=true]{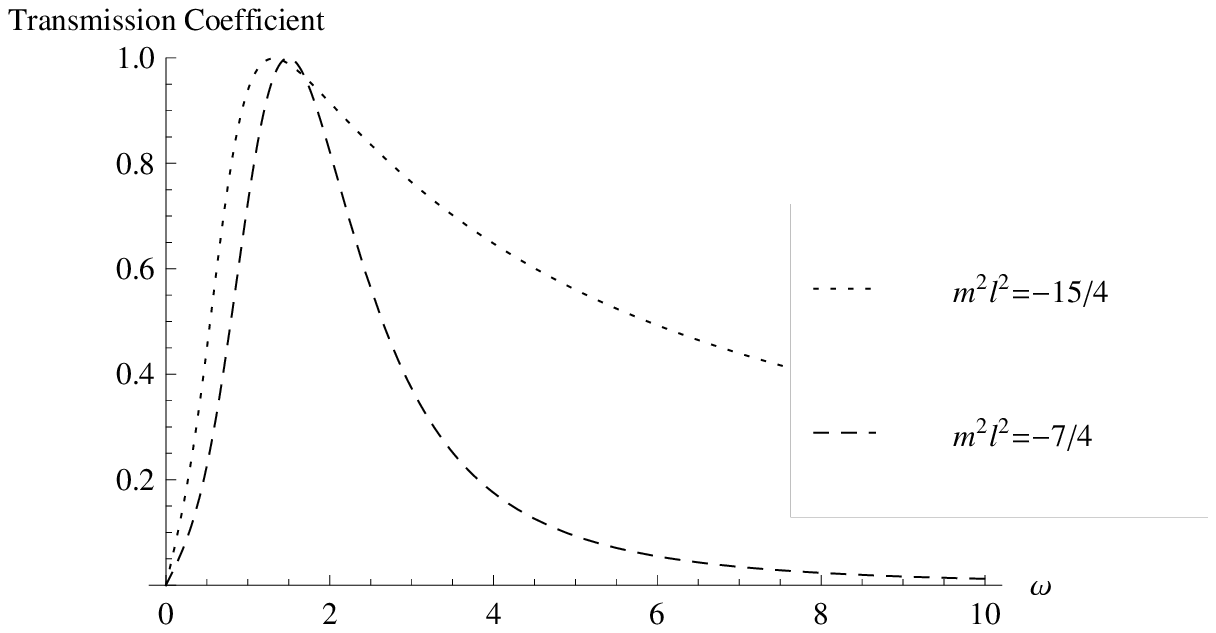}
\caption{Transmission coefficient v/s $\omega$; $d=5$,
$m^2l^2=-15/4,-7/4$, $l=1$, $h=-1$, $p=0.5$ and $\xi=0$.}
\label{TransmissionCoefficientCSBH5dmA}
\end{figure}

\begin{figure}
\includegraphics[width=4.0in,angle=0,clip=true]{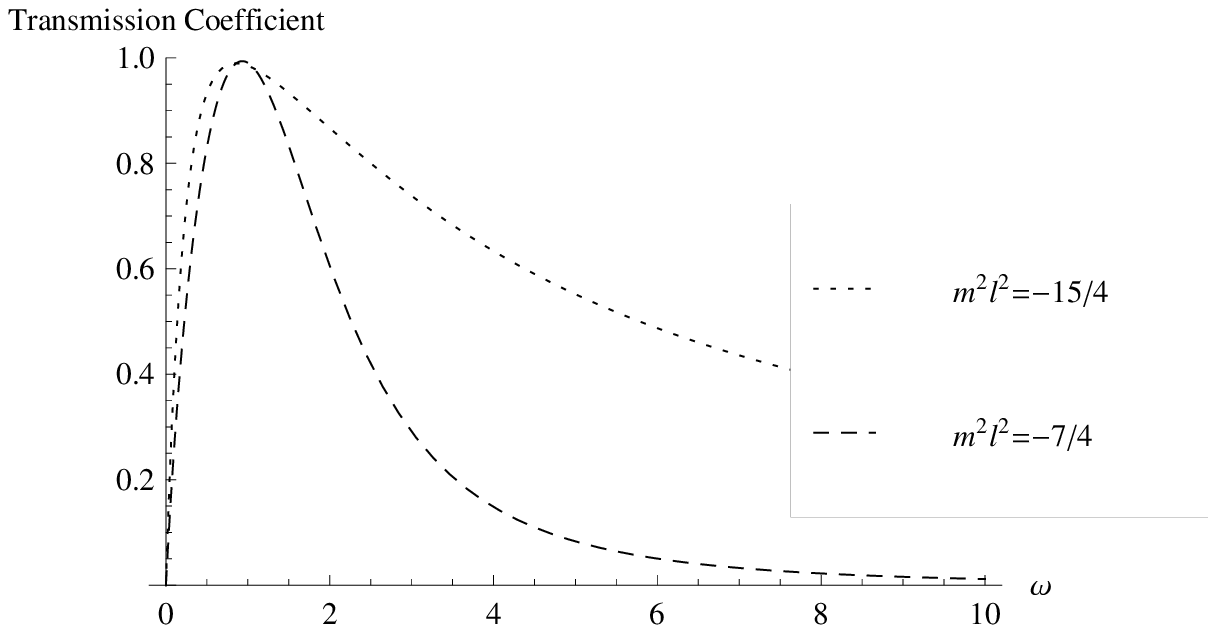}
\caption{Transmission coefficient v/s $\omega$; $d=5$,
$m^2l^2=-15/4,-7/4$, $l=1$, $h=-1$, $p=2$ and $\xi=0$.}
\label{TransmissionCoefficientCSBH5dmB}
\end{figure}

\begin{figure}
\includegraphics[width=4.0in,angle=0,clip=true]{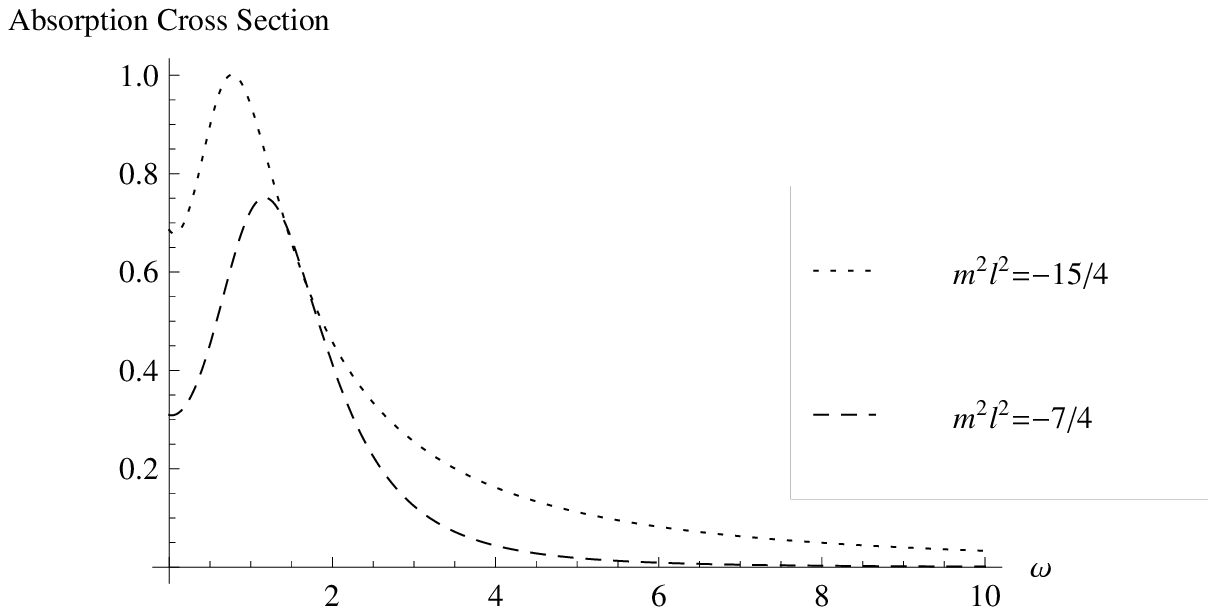}
\caption{Absorption Cross Section v/s $\omega$; $d=5$,
$m^2l^2=-15/4,-7/4$, $l=1$, $h=-1$, $p=0.5$ and $\xi=0$.}
\label{AbsorptionCrossSectionCSBH5dmA}
\end{figure}

\begin{figure}
\includegraphics[width=4.0in,angle=0,clip=true]{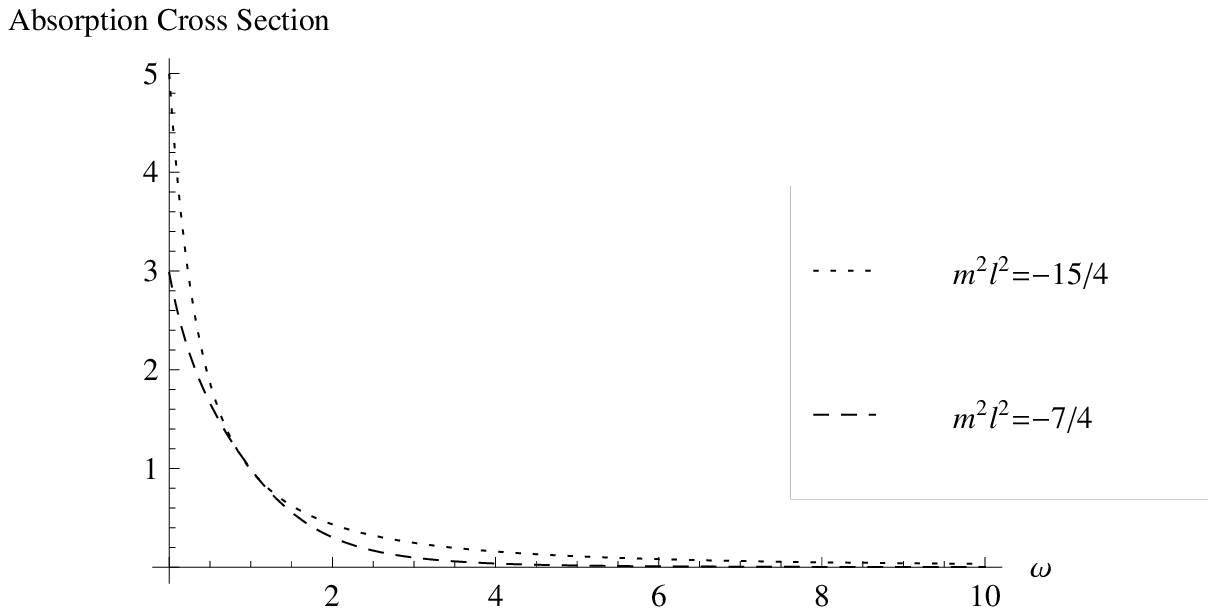}
\caption{Absorption Cross Section v/s $\omega$; $d=5$,
$m^2l^2=-15/4,-7/4$, $l=1$, $h=-1$, $p=2$ and $\xi=0$.}
\label{AbsorptionCrossSectionCSBH5dmB}
\end{figure}

  We consider next  without loss of generality, $m^2l^2=-15/4$, $l=1$, $p=0.5,2$ and we fix $h=-1$,
 and then we  analyze the behaviour of the coefficients in five dimensions for various values of $\xi$.
 Our results  for $\xi=0,1,2,2.5$ are shown  in
 Figs.~(\ref{ReflectionCoefficientCSBH5dA}, \ref{ReflectionCoefficientCSBH5dB}),  (\ref{TransmissionCoefficientCSBH5dA},
  \ref{TransmissionCoefficientCSBH5dB}), and (\ref{AbsorptionCrossSectionCSBH5dA}, \ref{AbsorptionCrossSectionCSBH5dB}), for the reflection,
 transmission coefficients and the greybody factors, respectively.
  We found, in the zero-frequency limit that there is a range of values of $\xi$ that
 contribute to the greybody factor, in contrast to the  case analyzed by Das, Gibbons and Mathur \cite{Das:1996we},
   where in the zero-frequency limit, only the mode with lowest
 angular momentum contributes to the absorption cross section.
 Also we plot in
Figs.~(\ref{ReflectionCoefficientCSBH5dmA},
\ref{ReflectionCoefficientCSBH5dmB}),
(\ref{TransmissionCoefficientCSBH5dmA},
\ref{TransmissionCoefficientCSBH5dmB}) and
(\ref{AbsorptionCrossSectionCSBH5dmA},
\ref{AbsorptionCrossSectionCSBH5dmB}),  the mode with lowest
angular momentum $\xi=0$  for  $d=5$, $m^2l^2=-15/4,-7/4$,
$p=0.5,2$, $l=1$ and $h=-1$, the reflection  the transmission
coefficient and the greybody factor  respectively.

We observed in the low frequency limit, that the reflection and
transmission coefficients show a minimum and a maximum. Therefore,
the coefficients have two branches in the reflection case,
decreasing for low frequencies and then increasing. In the
transmission case the behaviour is opposite, they are increasing
and then decreasing, in such way as $\Re+\mathfrak{U}=1$ in all
cases. An interesting result is the existence of one optimal
frequency to transfer energy out of the bulk.










\section{Conclusions}

Chern-Simons black holes are very interesting static solutions of
Gravity theories which asymptotically approach spacetimes of
constant negative curvature (AdS spacetimes). They can be
considered as generalizations of the (2+1)-dimensional black holes
in higher-dimensional Gravity theories containing higher powers of
curvature. The Chern-Simons black holes of spherical topology have
the same causal structure as the BTZ black holes and these
solutions have a thermodynamical behavior which is unique among
all possible black holes in competing Lanczos-Lovelock theories
with the same asymptotics. The specific heat of these black holes
is positive and therefore they can always reach thermal
equilibrium with their surroundings and hence, are stable against
thermal fluctuations.

These theories of high curvature also admit solutions which
represent black objects of other topologies, whose singularity is
surrounded  by horizons of non-spherical topology.  In the case of
hyperbolic topology, introducing a discrete symmetry group and
making the right identifications the resulting black holes
resemble the topological black holes. The Chern-Simons black holes
with hyperbolic topology have a single horizon, and the
temperature is a linear function of the horizon. It is interesting
one to study the possibility of the existence of a phase
transition as it happens in the case of the topological black
holes~\cite{Koutsoumbas:2008yq}.

In this work we calculated the QNMs of scalar perturbations of the
Chern-Simons black holes with hyperbolic topology. We found that
the QNMs depend on the highest power of curvature present in the
Lanczos-Lovelock theories. We also calculated the mass and area
spectrum of these black holes. We found that there is a strong
dependence of the quantization conditions on the underlying
geometry resulting in making these conditions not evenly spaced.

We also  computed analytically the greybody factors for
Chern-Simons black holes in  $d$-dimensions. We made a numerical
analysis of the behaviour
 of the reflection and the transmission
coefficients and the greybody factors in the low frequency limit
and we found that there is a range of  modes which contributes to
the absorption cross section.



\section*{Acknowledgments}

We thank Christos Charmousis, Olivera Miskovic, Julio Oliva,
George Siopsis,
 Petros Skamagoulis and Ricardo Troncoso for stimulating discussions. P.G. was
supported by Direcci\'{o}n de Estudios Avanzados PUCV.


\appendix

\end{document}